\documentclass{aa}
\usepackage{graphics}
\usepackage{amssymb}
\usepackage{natbib}
\usepackage{threeparttable}
\bibpunct{(}{)}{;}{a}{}{,}  

\newcommand{\eqb}{\begin{eqnarray}}
\newcommand{\eqe}{\end{eqnarray}}
\newcommand{\sth}{\sigma_{\rm T}}
\newcommand{\sgg}{\sigma_{\gamma \gamma}}
\newcommand{\gmn}{\gamma_{\rm min}}
\newcommand{\gmx}{\gamma_{\rm max}}
\newcommand{\gcr}{\gamma_{\rm br}}

\newcommand{\me}{m_{\rm e}}

\newcommand{\lB}{\ell_{\rm B}}

\newcommand{\tcr}{t_{\rm cr}}

\newcommand{\Bcr}{B_{\rm cr}}
\newcommand{\teesc}{{t}_{\rm e, esc}}
\newcommand{\tgesc}{{t}_{\rm \gamma, esc}}

\newcommand{\nga}{n_{\gamma}}
\newcommand{\ns}{n_{\rm s}}
\newcommand{\next}{n_{\rm ex}}
\newcommand{\nel}{n_{\rm e}}
\newcommand{\emn}{\epsilon_{\rm min}}
\newcommand{\emx}{\epsilon_{\rm max}}

\newcommand{\eq}{\epsilon_{\rm q}}
\newcommand{\ecr}{\epsilon_{\rm br}}
\newcommand{\xmx}{x_{\rm max}}
\newcommand{\lcr}{\ell_{\rm \gamma, cr}}
\newcommand{\linj}{\ell_{\rm inj}}
\newcommand{\leinj}{\ell^{\rm e}_{\rm inj}}

\begin{document}

\title{Spontaneously quenched gamma-ray spectra from compact sources}
\author{M. Petropoulou, D. Arfani \and A. Mastichiadis}
\institute{Department of Physics, University of Athens, Panepistimiopolis, GR 15783 Zografos, Greece}
\date{Received.../Accepted...}


\abstract{}
{We study a mechanism for producing intrinsic broken power-law $\gamma$-ray
spectra in compact sources. This is based on the principles of automatic photon quenching, according to which,
$\gamma$-rays are being absorbed on spontaneously produced soft photons, whenever
the injected luminosity in $\gamma$-rays lies above a certain critical value.}
{We derive an analytical expression for the critical $\gamma$-ray compactness in the case of
power-law injection. For the case where automatic photon quenching is relevant, we 
calculate analytically the emergent steady-state $\gamma$-ray spectra. We perform also numerical calculations
in order to back up our analytical results.}
{We show that a spontaneously quenched power-law $\gamma$-ray spectrum obtains
a photon index $3 \Gamma/2$, where $\Gamma$ is the photon index of the power-law at
injection. Thus, large spectral breaks of the $\gamma$-ray photon spectrum, e.g. $\Delta \Gamma \gtrsim 1$,
can be obtained by this mechanism. We also discuss additional features
of this mechanism that can be tested observationally. Finally, we fit the multiwavelength
spectrum of a newly discovered blazar (PKS 0447-439) by using such parameters, as to
explain the break in the $\gamma$-ray spectrum by means of spontaneous photon quenching, under
the assumption that its redshift lies in the range $0.1<z<0.24$.}
{}


\titlerunning{}
\authorrunning{Petropoulou, Arfani \& Mastichiadis}
\maketitle

\section{Introduction}

The production and radiation transfer of high-energy $\gamma$-rays
is a physical problem that has attained a lot of attention over the last
forty years since it can be applied on compact high-energy emitting astrophysical
sources, such as Active Galactic Nuclei (AGN) and Gamma-Ray Bursts. 
Photon-photon absorption, in particular, turns out to be a significant
physical process in compact X-ray and $\gamma$-ray emitting sources
that results in electromagnetic (EM) cascades (e.g. \cite{jelley66, herterich74}).
The effects of EM cascades
can be studied within either a linear or non-linear framework. 
In the first approach (e.g. \cite{protheroe86}),
the number density of target photons is assumed to be fixed, whereas in the second one, 
the produced electron/positron pairs produce photons, which on their turn serve  as targets
for photon-photon absorption \citep{kazanas84,zdiarski85,svensson87}.
The first analytical studies of EM cascades were then followed by 
numerical works, which aimed  at computing time-dependent
solutions to the kinetic equations of electrons and photons
 taking photon-photon annihilation
into account \citep{coppi92, mastkirk95, stern95, boettcherchiang02}. 
These algorithms are now commonly used in source modelling 
\citep{mastkirk97, kataoka00, konopelko03, katarz05}. 

However, intrinsically non-linear effects in EM cascades 
initiated by photon-photon absorption  have only recently 
gained attention. First, \cite{SK07} -- from this point on SK07 --
studied the case where no soft (target) photons are present
in a source. They investigated the necessary conditions 
under which $\gamma$-ray photons can cause runaway
pair production and found 
that these conditions can be summarized only in  a single quantity, the `critical'
$\gamma$-ray compactness. This can be considered as an upper limit of the $\gamma$-ray compactness,
that depends on parameters such as the magnetic field strength and the size of the source.
If the injection compactness of very high energy  (VHE)
 photons ($\gtrsim 0.1$ TeV) is larger than the critical one, the following non-linear loop
is self-sustained: $\gamma$-ray photons are absorbed on soft photons emitted by the
produced pairs through synchrotron radiation.

The work of SK07 was then expanded by \cite{PM11} -- henceforth PM11 -- mainly by taking into account
continuous energy losses of the produced pairs. The non-linear loop of processes 
called `automatic photon quenching' by SK07 can be the core of other more complex ones.
In particular, \cite{PM12} or just PM12b from this point on, have attributed the production
of VHE $\gamma$-rays to synchrotron emission from secondaries produced
in charged pion decay, while pions were the result of  photohadronic interactions between
relativistic protons and soft photons. PM12b have shown that the system of protons and photons
resembles that of a prey-predator one, 
whenever automatic photon quenching operates, and it
shows interesting variability patterns, such as limit cycles\footnote{Similar results are presented
in \cite{mastetal05} but they are caused by a different intrinsic non-linear process known as the `PPS-loop' \citep{kirkmast92}.}.

In the present work we continue the exploration of spontaneous photon quenching by 
studying the case of power-law $\gamma$-ray injection in the source and, in that sense, it
can be considered as a continuation of the aforementioned works. There were two main motivations of our present study: (i) $\gamma$-ray spectra 
emitted by a power-law 
distribution of relativistic particles through some radiation 
mechanism, e.g. synchrotron radiation and inverse Compton scattering,
can be modelled by a power-law, at least partially and 
(ii) if spontaneous photon absorption affects part of 
the $\gamma$-ray injection spectrum, spectral breaks are produced;
we believe that this requires further investigation, as it
is an \textit{intrinsic} mechanism for producing breaks in a $\gamma$-ray spectrum and
it could be of relevance to recent results regarding 
 $\gamma$-ray emitting blazars.

The present paper is structured as follows: in  section 2 we 
derive an analytical expression for the critical $\gamma$-ray compactness 
in the case of power-law injection using certain simplifying assumptions, while
we comment also on the validity range of our result. In the case where spontaneous photon
quenching becomes relevant, we show that a break at the steady-state $\gamma$-ray spectrum appears
and we further calculate analytically the expected  spectral change. In section 3 we derive numerically the
critical compactness for a wide range of parameter values and examine the effects 
that a primary soft photon component in the source would have on $\gamma$-ray absorption. Possible implications
of spontaneous absorption on $\gamma$-ray emitting blazars are presented in section 4. We also
present a list of observationally tested characteristics that a spontaneously quenched source would, in principle, show.
In the same section we further show that the spectral energy distribution (SED) of the newly discovered blazar PKS 0477-439 can 
be explained
within the framework of automatic quenching. For 
the required transformations
between the reference systems of the blazar and the observer 
 we have adopted a cosmology 
with $\Omega_{\rm m}=0.3$, $\Omega_{\Lambda}=0.7$ and 
$H_0=70$ km s$^{-1}$ Mpc$^{-1}$.

\section{Analytical approach}
We consider a spherical region of radius $R$ containing a magnetic field 
of strength $B$. We assume that $\gamma$-rays are being produced in this
volume by some non-thermal emission process, e.g. proton synchrotron radiation. In our
analysis however, the $\gamma$-ray production mechanism remains unspecified, since its
exact nature does not play a role in the derivation of our results. Furthermore, $\gamma$-rays
are being injected with a luminosity $L_\gamma^{\rm inj}$  that
is related to the injected $\gamma-$ray compactness as
\eqb 
\label{lgg}
\linj={{L_\gamma^{\rm inj}\sth}\over
{4\pi R\me c^3}},
\eqe
where $\sigma_T$ is the Thomson cross section. 
Without any substantial soft photon population 
inside the source, the $\gamma$-rays
will escape without any attenuation in one crossing time.
However, as SK07 and PM11 showed, the injected 
$\gamma$-ray compactness cannot 
become arbitrarily high because, if a critical value is reached, 
the following loop starts operating: (i) 
Gamma-rays pair-produce on soft photons, which can be
initially arbitrarily low inside the source; (ii) 
the produced electron-positron pairs are highly relativistic, since they are created with
approximately half the energy of the initial $\gamma$-ray photon, and cool mainly by synchrotron radiation,
 thus acting as a source of soft photons; (iii) 
the emitted synchrotron photons have lower energy when compared to the $\gamma$-ray photons and 
 serve as targets for more $\gamma \gamma$ interactions.

There are two conditions that should
be satisfied simultaneously for this 
network to occur: 
\begin{enumerate}
 \item \textit{Feedback} criterion \\
       This is related to the energy threshold condition for photon-photon
absorption and it requires that the synchrotron photons emitted from the pairs
have sufficient energy to pair-produce on the $\gamma$-rays. \\
\item \textit{Marginal stability} criterion \\
 This is related to the optical depth for photon-photon absorption, which
must be above unity in order to establish the growth of the instability.
\end{enumerate} 
By making suitable simplifying assumptions, one can derive 
an analytic relation for the first condition -- see also SK07. Thus, 
combining (i) the threshold condition for $\gamma \gamma$ absorption
$\epsilon x=2$, where $\epsilon$ and $x$ are the $\gamma$-ray and soft photon energies in units of $\me c^2$ respectively  -- this normalization will be used
for all photon energies throughout the text unless stated otherwise --
(ii) the fact that there is equipartition of energy among
the created electron-positron pairs $\gamma_e=\gamma=\epsilon/2$
and (iii) the assumption that the required
soft photons are the synchrotron photons that the electrons/positrons radiate,
i.e., $x_{\rm s}=b\gamma^2$
where $b=B/B_{\rm crit}$ and $B_{\rm crit}=(\me^2 c^3)/(e\hbar)
\simeq 4.4\times 10^{13}$~G, one finds the following relation
\eqb
\epsilon_{\rm q} = \frac{2}{b^{1/3}}
\label{ecrit}
\eqe
that defines, for a certain magnetic field strength, the $\gamma$-ray photon energy
above which automatic photon quenching becomes relevant. 

In what follows, we will concentrate on the second condition, since our aim is to determine the value 
of the injected $\gamma$-ray compactness that ensures the growth of the instability.
The corresponding calculations in the case of monoenergetic $\gamma$-ray injection can be found in SK07 and PM11.
Here we focus on the more astrophysically relevant case of a power-law $\gamma$-ray injection. In order to treat this problem analytically we 
`discretize' the power-law of $\gamma$-rays. In particular, we begin
by calculating the critical compactness in the case where two monoenergetic $\gamma$-rays are injected. We repeat
the calculation for the injection of three monoenergetic $\gamma$-rays and, finally, we generalize our result for N
monoenergetic injection functions. 
Furthermore, our treatment is built upon the following assumptions \& approximations:
\begin{enumerate}
 \item Only two physical processes are taken into account, i.e. photon-photon absorption
and synchrotron radiation of the produced pairs. Inverse Compton scattering of pairs on the 
synchrotron produced photons can be safely neglected because of the strong magnetic field, that is 
typically required for the automatic photon quenching loop to function, and the large\footnote{For a $100$ GeV $\gamma$-ray photon, 
the dimensionless photon energy is $\epsilon=2\times10^5$ and the produced pairs have $\gamma \approx \epsilon/2 = 10^5$.} Lorentz factors of the produced
pairs. 
\item Only the equations describing the evolution of $\gamma$-rays and synchrotron (soft) photons are taken into
account. The equation for the pairs is neglected, since these have synchrotron cooling timescales much smaller than
the crossing time of the source. Thus, all the injected energy into pairs is transformed into synchrotron radiation.
\item Synchrotron emissivity is approximated by a $\delta$-function, i.e.,
$j_s(x)=j_0 \delta(x-x_{\rm s})$, where $x_{\rm s}=b\gamma^2$ is
 the synchrotron critical energy.
\item The synchrotron energy losses of pairs are treated as `catastrophic' escape from the considered energy range. In other
words, an electron with  Lorentz factor $\gamma$ loses its energy by radiating synchrotron photons at energy $b\epsilon^2/4$.
\item The cross section of photon-photon absorption (in units of $\sth$) is approximated as  
\eqb
\sgg \approx  \sigma_0 \frac{H(x\epsilon-2)}{x\epsilon}, \ \sigma_0=0.652
\eqe
which is the same as the one given by \cite{coppibland90} apart from the logarithmic term $\ln(x)$. 
\end{enumerate}

\subsection{Marginal stability criterion for injection of two monoenergetic $\gamma$-rays}
Let us assume that $\gamma$-rays are being injected into the source 
at energies $\epsilon_1$ and $\epsilon_2$ ($\epsilon_1 < \epsilon_2$) with compactnesses $\linj^{(i)}$ where $i=1,2$.
We use such parameters in order to ensure that the $\gamma$-ray photon with the minimum energy
satisfies the feedback criterion, i.e. $\epsilon_1>\eq=2/b^{1/3}$. Then, all higher energy $\gamma$-ray photons
also satisfy the feedback criterion and the corresponding emitted synchrotron photons have energies  
$x_{\textrm {s}, i}=b\epsilon_i^2/4$.
Gamma-ray photons with energy $\epsilon_i$ can, therefore, be absorbed on both soft photon
distributions because the energy threshold criterion is satisfied for all the four possible photon-photon
interactions,i.e. $\epsilon_i x_{\textrm {s},j}>2$ for $i,j=1,2$. We note also that
we do not consider absorption of $\gamma$-rays on less energetic $\gamma$-rays.
Moreover, the number densities of $\gamma$-ray photons and of the corresponding soft photons 
are denoted as $\nga^{(i)}$ and $\ns^{(i)}$ respectively, where
the symbols imply the relations 
$\nga^{(i)}\equiv \nga(\epsilon)\delta(\epsilon-\epsilon_i)$ and $\ns^{(j)}\equiv \ns(x)\delta(x-x_{\textrm{s},j})$; 
the densities refer to the number
of photons contained in a volume element $\sth R$. In other
words, if $\hat{n}$ expresses the number of photons
per erg per cm$^3$, then $n = \hat{n}(\sth R)(\me c^2)$.
The dimensionless photon number densities are also related to the compactnesses through
the relation
\eqb
\linj^{(i)}=\frac{\nga^{(i)}\epsilon_i^2}{3}, \ i=1,2
\eqe
for discrete monoenergetic injection\footnote{For a continuous power-law injection of photons
the relation between the differential photon number density and compactness is
 $\linj(\epsilon)=\nga(\epsilon)\epsilon/3$, while the total injection compactness
is calculated by $\linj=\int\textrm{d}\epsilon \ \linj(\epsilon)$}.
Using the notations introduced above along with the assumptions (1)-(5) the system can be described by 
four equations 
\eqb
\label{4eqa}
\frac{d\nga^{(i)}}{d \tau} +\nga^{(i)}  & = &  \textit{Q}_{\rm inj}^{(i)}+\textit{L}_{\gamma\gamma}^{(ij)}, \ i,j=1,2\\
\frac{d \ns^{(i)}}{d\tau} +\ns^{(i)} &=& \textit{Q}_{\gamma\gamma}^{(ij)}, \ i,j=1,2 
\label{4eqb}
\eqe
where time is normalized with respect to the photon crossing/escape time from 
the source, i.e. $\tau=ct/R$ and 
the operators  $\textit{L}$ and $\textit{Q}$ denote losses and injection respectively; 
the loss term in eq.~(\ref{4eqb}) due to photon-photon absorption is omitted since it is negligible.
We note that number densities and rates are equivalent in the dimensionless form of the above equations.
The explicit expressions of the operators in eqs.~(\ref{4eqa}) and (\ref{4eqb}) are
\eqb
\textit{L}_{\gamma \gamma}^{(ij)} & = & - \nga^{(i)}\int\textrm{d}x \ \sgg(x\epsilon_i) \ns(x)\delta(x-x_{\textrm{s},j})= \nonumber \\
 & = & - \frac{\sigma_0}{\epsilon_i x_{\textrm{s},j}}\nga^{(i)}\ns^{(j)} \\
\textit{Q}_{\gamma \gamma}^{(ij)} & = & \textit{Q}_{\gamma\gamma}^{0}\nga^{(i)}\ns^{(j)}\\
\label{qinj}
\textit{Q}_{\rm inj}^{(i)} & = & \frac{3\linj^{(i)}}{\epsilon_i^2},
\eqe
where the normalization constant $\textit{Q}_{\gamma\gamma}^{0}$ is calculated  by equating the total $\gamma$-ray energy loss rate with
the total energy injection rate into soft photons, i.e.
 $-\int\textrm{d}\epsilon \ \epsilon \textit{L}_{\gamma\gamma} =  \int \textrm{d}x \ x \textit{Q}_{\gamma\gamma}$, and it is given by
%
\eqb
\textit{Q}^{0}_{\gamma\gamma} =  \frac{\sigma_0}{x_{\textrm{s},j}^2}.
\eqe
We note also that in the case of continuous power-law injection
 eq.~(\ref{qinj}) should be replaced by $\textit{Q}_{\rm inj}(\epsilon)=3\linj(\epsilon)/\epsilon$.
The trivial stationary solution of eqs.~(\ref{4eqa}) and (\ref{4eqb})
is $\left(\bar{n}_{\gamma}^{(1)},\bar{n}_{\gamma}^{(2)},0,0\right)$, where 
$\bar{n}_{\gamma}^{(i)}=\textit{Q}_{\rm inj}^{(i)}$ and it 
corresponds to the case where the injection rate
of $\gamma$-rays equals the photon escape rate from the source.
Following the methodology described in 
SK07 and PM11 we introduce perturbations to all photon number densities and
linearize the set of equations (\ref{4eqa})-(\ref{4eqb}) around
the trivial solution. The linearized system can be 
written in the form $dY/d\tau = \mathbf{A} Y$ where
\eqb
Y = \left( \begin{tabular}{c}
            $\nga^{'(1)}$ \\
            $\nga^{'(2)}$ \\
   $\ns^{'(1)}$ \\
  $\ns^{'(2)}$
            \end{tabular}
\right)
\eqe
is the vector of the perturbed number densities 
and $\mathbf{A}$ is the matrix of the linearized system of equations
\eqb
 \mathbf{A}  =  \left( \begin{tabular}{cccc} 
-1&0 &-$\frac{\sigma_0}{\epsilon_1x_{\textrm{s},1}} \bar{n}_{\gamma}^{(1)}$&  $-\frac{\sigma_0}{\epsilon_1 x_{\textrm{s},2}} \bar{n}_{\gamma}^{(1)}$ \\
0 & -1 & $-\frac{\sigma_0}{\epsilon_2 x_{\textrm{s},1}} \bar{n}_{\gamma}^{(2)}$ & $-\frac{\sigma_0}{\epsilon_2 x_{\textrm{s},2}} \bar{n}_{\gamma}^{(2)} $\\
0 & 0 & $-1 + \frac{\sigma_0}{x_{\textrm{s},1}^2} \bar{n}_{\gamma}^{(1)} $& $\frac{\sigma_0}{x_{\textrm{s},2}^2} \bar{n}_{\gamma}^{(1)} $\\
0 & 0 & $\frac{\sigma_0}{x_{\textrm{s},1}^2} \bar{n}_{\gamma}^{(2)}$ & $-1+\frac{\sigma_0}{x_{\textrm{s},2}^2} \bar{n}_{\gamma}^{(2)} $    
                                           \end{tabular}
\right).
\eqe 
In order to build a finite number of soft photons in the source, the perturbations must grow with time. This is ensured, if, at least one of the eigenvalues
 of matrix $\mathbf{A}$ is positive. After some algebraic manipulation we find that, indeed, one eigenvalue can become
 positive if the following
condition holds
\eqb
\frac{\textit{Q}_{\rm inj}^{(1)}}{\epsilon_1^4} + \frac{\textit{Q}_{\rm inj}^{(2)}}{\epsilon_2^4} \ge \frac{b^2}{16\sigma_0}.
\label{lcr_two}
\eqe
The same methodology can be applied in the case where $\gamma$-rays are being injected as a $\delta$- function at 
three energies $\epsilon_1<\epsilon_2<\epsilon_3$. This leads to an analogous critical condition
\eqb
\frac{\textit{Q}_{\rm inj}^{(1)}}{\epsilon_1^4} + \frac{\textit{Q}_{\rm inj}^{(2)}}{\epsilon_2^4} + \frac{\textit{Q}_{\rm inj}^{(3)}}{\epsilon_3^4} \ge \frac{b^2}{16\sigma_0}.
\label{lcr_three}
\eqe
\subsection{Critical compactness for power-law $\gamma$-ray injection}
The above can be generalized for the case of N monoenergetic $\gamma$ -rays with energies $\epsilon_1 < \epsilon_i <\epsilon_N$
in order to find the following marginal stability criterion
\eqb
\sum_{i=1}^{N}\frac{\textit{Q}_{\rm inj}^{(i)}}{\epsilon_i^4} \ge \frac{b^2}{16 \sigma_0}.
\label{lcr_N}
\eqe
We have verified that the above relation applies also to cases where
the feedback criterion is not satisfied for $\gamma$-rays of the minimum energy but for higher energy photons, i.e.
 $\epsilon_k=2/b^{1/3}$ with $k>1$, with only a slight modification: the summation starts
from $i=k$.  Moreover, if we assume that the injection rate of $\gamma$-ray photons can be modelled as a power-law, e.g.
\eqb
\textit{Q}_{\rm inj}^{(i)} = \textit{Q}_0 \left(\frac{\epsilon_i}{\epsilon_1}\right)^{-\Gamma}, \ i=1,\dots, N
\eqe
the criterion of eq.~(\ref{lcr_N}) takes the form
\eqb
\textit{Q}_0 \sum_{i=1}^{N} \epsilon_i^{-\Gamma-4} \ge \frac{b^2}{16 \sigma_0 \epsilon_1^{\Gamma}}.
\label{lcr_N_a}
\eqe
 If $N\rightarrow \infty$ and $(\epsilon_{i+1}-\epsilon_i)\rightarrow 0$, 
the discrete sum of eq.~(\ref{lcr_N_a}) can be trasformed into an integral which leads to 
\eqb
\textit{Q}_0 \ge \frac{b^2\epsilon_1^{-\Gamma}}{16\sigma_0}\frac{\Gamma+3}{\epsilon_1^{-\Gamma-3}-\epsilon_N^{-\Gamma-3}}.
\eqe
Finally, if we replace the normalization constant $\textit{Q}_0$ by the integrated $\gamma$-ray compactness 
over all photon energies (see also comment on eq.~(\ref{qinj})) using  
\eqb
\linj=\frac{1}{3}\int_{\epsilon_1}^{\epsilon_N} \textrm{d}\epsilon \ \textit{Q}_0 \epsilon \left(\frac{\epsilon}{\epsilon_1}\right)^{-\Gamma}
\eqe 
we find that $\linj\ge \lcr$ where $\lcr$ is the critical compactness for a power-law $\gamma$-ray injection. This has a rather compact form
\eqb
\lcr = \frac{b^2}{48\sigma_0}\left\{
 \begin{array}{ll}
\frac{(\Gamma+3)\left(\emn^{-\Gamma+2}-\emx^{-\Gamma+2}\right)}{(\Gamma-2)\left(\epsilon_{\rm M}^{-\Gamma-3}-\emx^{-\Gamma-3}\right)},&\textrm{for}\ \Gamma \ne 2 \\
\phantom{} & \phantom{}\\
 \frac{(\Gamma+3)}{\left(\epsilon_{\rm M}^{-\Gamma-3}-\emx^{-\Gamma-3}\right)}\ln\left(\frac{\emx}{\emn}\right), &   \textrm{for} \ \Gamma = 2,              
 \end{array}
\right.
\label{lcr}
\eqe
where $\emn\equiv \epsilon_1$, $\emx\equiv \epsilon_N$ and $\epsilon_{\rm M}=\max[\emn,\eq]$ -- for the definition of $\eq$ see 
eq.~(\ref{ecrit}).
 
\begin{figure}
\begin{tabular}{l}
 \includegraphics[width=8.2cm, height=7cm]{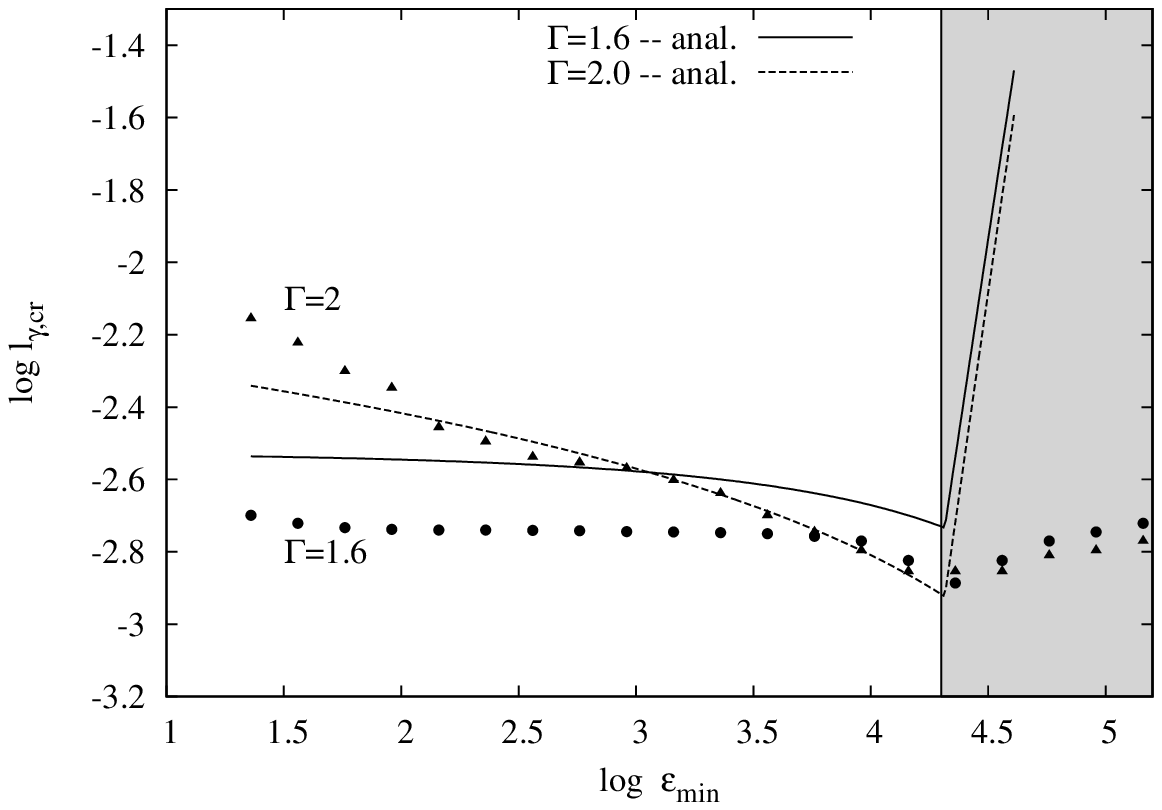}\\
\includegraphics[width=8.2cm, height=7cm]{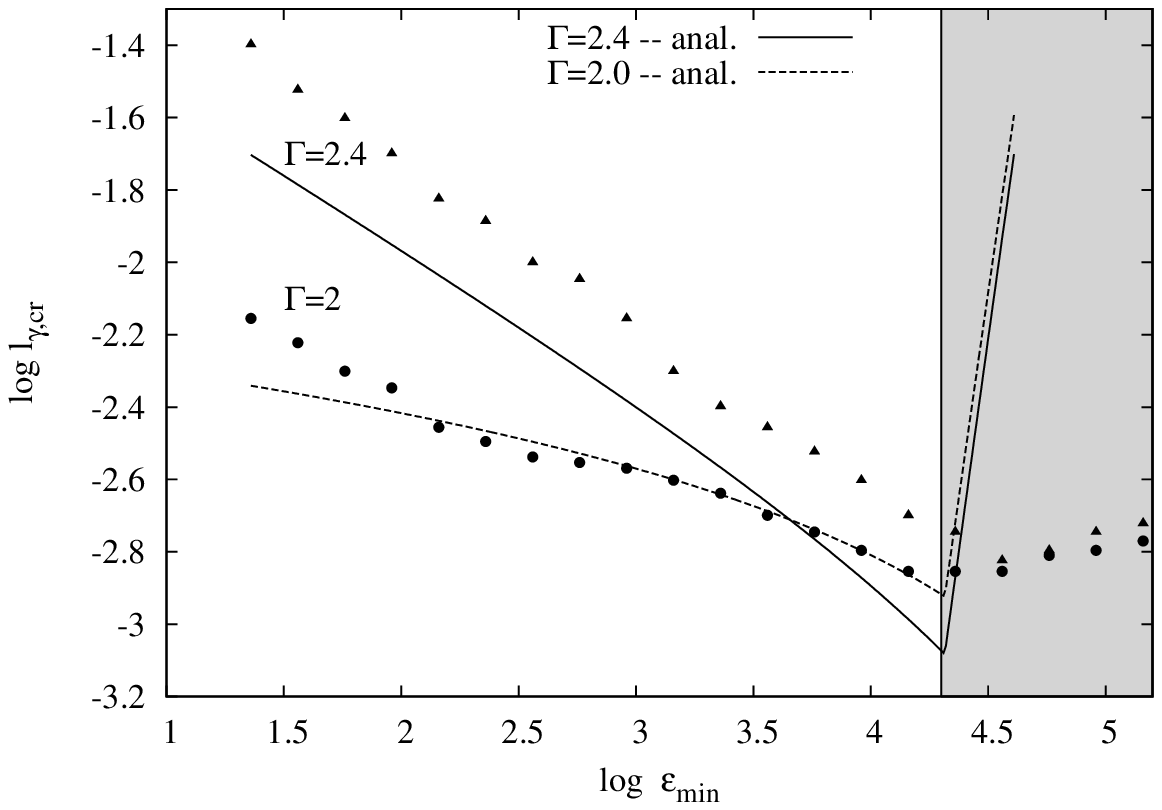}
\end{tabular}
\caption{Top panel: critical compactness $\lcr$ as a function of the minimum energy of the $\gamma$-ray
spectrum $\emn$ for $\Gamma=1.6$ (solid line) and $\Gamma=2$ (dashed line).
The numerically derived values for the two cases are shown with circles and triangles respectively.
 Bottom panel: same as in top panel except for different photon indices, which are marked on the plot.
Other parameters used are: $\emx=2.3\times10^5$ (in $\me c^2$ units), $B=40$ G and $R=3\times10^{16}$ cm.
In both panels the grey area denotes the region where $\emn>\eq$ -- for the definition of $\eq$
see text.}
\label{lcr-fig}
\end{figure}

We were able to derive an analytical and rather simple expression of the critical compactness in the 
case of power-law injection at the cost, however, of a series of approximations/assumptions that may limit 
the validity range of our result. It is reasonable therefore, before closing the present section
to check the range of validity of eq.~(\ref{lcr}). For this, we made a comparison between this expression and the numerically derived
values\footnote{For more details
on the numerical code used, see section 3.}, 
which is exemplified in Fig.~\ref{lcr-fig}. 
Both panels  show the dependence of $\lcr$ on $\emn$ for two different pairs of photon indices marked on the plots.
Lines and symbols are used for plotting the analytically and numerically derived values respectively, while 
different types of lines/symbols correspond to different values of the photon index.
For values of $\emn$ below $\eq$, which for the values used in this example equals $2\times 10^4$,
 we find a good agreement between our analytical and numerical results, apart from
some numerical factor of $\simeq 2$; notice that the dependence of $\lcr$ given by eq.~(\ref{lcr}) on $\emn$ coincides
with the one determined numerically. However, the analytical solution of $\lcr$ fails in the energy range of $\emn \ge\eq$, since, in this
regime, approximation (4) listed at the beginning of  section 2 proves to be crude.
%

\subsection{Steady state $\gamma$-ray quenched spectrum}
Assuming that the injected $\gamma$-ray compactness exceeds 
the critical value derived in the previous
section, we search for steady-state solutions of the system.
There is a main difference between the analytical approach that we follow in this section
and the one presented in  section 2.1: here we take into account the cooling of pairs
due to synchrotron radiation, i.e. synchrotron losses are treated as a continuous energy
loss  mechanism. For this reason, in addition to the equations of $\gamma$-rays and 
soft photons, we include in our analysis a third equation for electron/positron pairs or simply electrons from this point on.
Dropping the time derivatives, the kinetic equations for $\gamma$-rays,
soft photons and electrons are written respectively
as
\eqb
 \label{gammaray} \nga(\epsilon)= \textit{Q}^{\gamma}_{\rm inj}(\epsilon)+\textit{L}^{\gamma}_{\gamma \gamma}(\epsilon)
  \eqe
\eqb \label{soft} \ns(x) = \textit{Q}^{\rm s}_{\rm syn}(x)
  \eqe
and 
\eqb \label{elec} \nel(\gamma)=\textit{Q}^{\rm e}_{\gamma \gamma}(\gamma)+\textit{L}^{\rm e}_{\rm syn}(\gamma),
  \eqe
where $\nel$ is the dimensionless 
 electron
  number density and an electron escape timescale $t_{\rm e, esc}=\tcr$ was assumed. 
All normalizations
and approximations -- apart from the fourth in our list -- are the same as in section 2.1.
The
synchrotron emissivity is approximated by a $\delta$-function -- see section 2.1. 
 We note that in what follows we can safely neglect electron escape from the source,
since, for magnetic field strengths relevant to automatic quenching, synchrotron cooling is the dominant term in the electron equation; thus,
the left hand side of eq.~(\ref{elec}) is essentially equal to zero. 
Then, the injection and loss operators take the following forms:
\eqb
  \label{injection}
   \cal{Q}^{\gamma}_{\rm inj} & = & Q_0 \epsilon^{-\Gamma}H(\epsilon-\emn)H(\emx-\epsilon) \\
  \cal{Q}^{\rm s}_{\rm syn} & = & \alpha_1 x^{-1/2} \nel \left(\sqrt{\frac{x}{b}}\right) \\
\cal{Q}^{\rm e} _{\gamma \gamma} & =& -4 \cal{L}^{\gamma}_{\gamma \gamma},
\eqe
where the form of $\cal{Q}^{\rm s}_{\rm syn}$ implies the use of a $\delta$-function for the synchrotron emissivity (see e.g. \cite{mastkirk95}) and
$\alpha_1= (2/3) \lB b^{-3/2}$. The loss operators are given by
\eqb
  \label{loss}
  \cal{L}^{\gamma}_{\gamma \gamma} & = & -\nga(\epsilon)\int_0^{\xmx} \ \textrm{d} x \ \sgg(x\epsilon) \ns(x) \\ 
  \cal{L}^{\rm e}_{\rm syn} & = & +\alpha_2\frac{\partial}{\partial
  \gamma}\left(\gamma^2 \nel \right)  
\eqe
where  $\xmx=b\gmx^2=b\emx^2/4$ is the maximum energy of the produced soft photon distribution\footnote{As a minimal condition, we assume that
the energy threshold criterion for automatic photon quenching is satisfied at least from the maximum energy $\gamma$-rays, i.e. $b\emx^3>8$.},
$\alpha_2=4\lB/3$ with $\lB=\sth R U_{\rm B}/ \me c^2$ being the `magnetic compactness' and $\sgg$ is the dimensionless (in units of $\sth$) cross section
for photon-photon absorption (see point (5) of section 2). 
Within this approximation eq. (\ref{loss}) takes the form 
\eqb
\cal{L}^{\gamma}_{\gamma \gamma} & = & -\nga(\epsilon)\int_{2/\epsilon}^{\xmx} \ \textrm{d} x \ \sigma_0 \frac{\ns(x)}{x\epsilon}
\label{Lgg}
\eqe
which further implies that $\gamma$-rays with energy $\epsilon < \ecr \equiv 2/\xmx$ cannot be absorbed. Thus, 
pairs with  $\gamma < \gcr \equiv \ecr/2$  cannot be produced, i.e. the injection term in eq.~(\ref{elec}) vanishes.
The above can be summarized by inserting the step function $H(\epsilon-\ecr)$ in the expression of $\cal{L}^{\gamma}_{\gamma \gamma}$ given by eq.~(\ref{Lgg}).
On the one hand, for $\gamma<\gcr$ the electron distribution has the trivial form  $\nel \propto \gamma^{-2}$. 
On the other hand, for $\gamma \ge \gcr$, 
the distribution is determined by synchrotron losses and pair injection. 

We now proceed to calculate the electron distribution
for $\gamma>\gcr$. After having inserted the above expressions for the operators into eqs.~(\ref{gammaray})-(\ref{elec}), we combine them
 into one non-linear integrodifferential equation, where the unknown function is $\nel(\gamma)$:
\eqb
-\alpha_2\frac{d}{d \gamma}\left( \gamma^2 \nel\right) =  4{Q}^{\gamma}_{\rm inj}(2\gamma)H(\gamma-\gcr) \frac{I(\nel;\gamma)}{1+I(\nel;\gamma)}, 
\label{integrodiff}
\eqe
where 
\eqb
I(\nel;\gamma) = \frac{\alpha_1\sigma_0}{2\gamma}\int_{1/\gamma}^{\xmx} \ \textrm{d}x\ \nel\left(\sqrt{\frac{x}{b}}\right)x^{-3/2}.
\eqe
Integration of eq.~(\ref{integrodiff}) leads to
\eqb
\alpha_2\left[\gamma'^2\nel(\gamma')\right]_{\gmx}^{\gamma}=\frac{Q_0}{2^{\Gamma-2}}\int_{\gamma_{\rm min}^{\rm eff}}^{\gmx}
\!\! \textrm{d}\gamma' \ \gamma'^{-\Gamma}\frac{I(\nel;\gamma')}
{1+I(\nel;\gamma')},
\label{integral}
\eqe
where we have used the notation $\left[Y\right]_{x_1}^{x_2}\equiv Y(x_2)-Y(x_1)$ and $\gamma_{\rm min}^{\rm eff}=\textrm{max}[\gamma,\gmn,\gcr]$.
If $\gcr>\gmn$ then $\gamma_{\rm min}^{\rm eff}=\gamma$, whereas if $\gcr<\gmn$ then $\gamma_{\rm min}^{\rm eff}$ can be equal either to $\gmn$
or $\gamma$. The condition $\gcr<\gmn$ corresponds to the physical case where the entire $\gamma$-ray power-law spectrum is affected by automatic 
quenching. Therefore, the steady-state $\gamma$-ray spectrum will show, in general, no break (see also numerical example in Fig.~8 of PM11). 
Since in the present work we are interested in producing broken power-law $\gamma$-ray spectra, we will examine only the case where $\gcr>\gmn$ and
 therefore $\gamma_{\rm min}^{\rm eff}=\gamma$. 
Since an exact solution of eq.~(\ref{integral}) cannot be obtained analytically, we assume 
a certain form for the solution we are searching for. A reasonable `guess' is that the electron distribution
is a power-law, i.e. $\nel=N_0 \gamma^{-p}$, where $N_0$ and $p$ are the  parameters to be defined. Inserting this function
into eq.~(\ref{integral}) we obtain
\eqb
\alpha_2 N_0\left[\gamma'^{-p+2}\right]_{\gmx}^{\gamma} = \frac{Q_0}{2^{\Gamma-2}}\int_{\gamma}^{\gmx}\!\!\textrm{d}\gamma' 
\frac{A(N_0,p) \gamma'^{-\Gamma+\beta}}{1+A(N_0,p)\gamma'^{\beta}},
\label{equation1}
\eqe
where $\beta=\frac{p-1}{2}$ and $A(N_0,p)=N_0 b^{p/2}\alpha_1\sigma_0/(p+1)$ is a function only of the unknown parameters for fixed values of the magnetic field and source 
size. 
We note also that we have neglected the contribution of terms calculated at the upper limit ($\xmx$) while performing the integral $I(\gamma')$, so
that the solution of the final integral appearing at the right hand side of eq.~(\ref{equation1}) can be expressed in closed form as
\eqb
\left[\frac{A\gamma'^{-\Gamma+\beta+1}}{-\Gamma+\beta+1} 
{_2F_{1}}\left(1,1+\frac{1-\Gamma}{\beta},2+\frac{1-\Gamma}{\beta};-A\gamma'^{\beta}\right)\right]_{\gamma}^{\gmx}\!\!\!\!,
\label{hypergeometric}
\eqe
where ${_2F_1}(a,b,c;z)$ is the Gauss hypergeometric function. Without making any further approximations the above equation
cannot be solved for $p$ and $N_0$ analytically. However, if 
the term $A\gamma'^{\beta}$ inside the integral is much larger than unity, this is, then,  greatly simplified  
 and eq.~(\ref{equation1}) takes the form
\eqb
-\alpha_2 N_0\left[\gamma'^{-p+2}\right]_{\gamma}^{\gmx} \approx \frac{Q_02^{-\Gamma+2}}{-\Gamma+1} 
\left[\gamma'^{-\Gamma+1}\right]_{\gamma}^{\gmx},
\label{equation2}
\eqe
which is satisfied for 
\eqb
p & = & \Gamma+1 \quad \textrm{and} \\
 N_0& = & \frac{2^{-\Gamma+2}Q_0}{\alpha_2 (\Gamma-1)}.
\eqe
The above solution is valid as long as $\Gamma \neq 1$. In the opposite case, one should search for more general forms
of the electron distribution, i.e. $\nel(\gamma)\propto N_0(\gamma)\gamma^{-p}$ and follow the same procedure. However,
for the purposes of the present work, the solution for $\Gamma\neq 1$ is sufficient.
Summarizing, the electron distribution is given by
\eqb
\nel(\gamma)\!\!\! &=&\!\! \!N_0\gcr^{-\Gamma-1}\left \{
\begin{array}{ll}
 \left(\frac{\gamma}{\gcr}\right)^{-2}, \ \textrm{for} \ \gmn \le \gamma\le \gcr, \\
\phantom{} \\  
 \left(\frac{\gamma}{\gcr}\right)^{-\Gamma-1}, \ \textrm{for} \ \gcr<\gamma\le \gmx
\end{array}
\right.
\label{elec-sol}
\eqe 
where we have demanded the solution to be continuous at $\gamma=\gcr$. 
Notice that the slope of the electron distribution above $\gcr$
is $\Gamma+1$, i.e. steeper than that at injection (see term $\gamma'^{-\Gamma}$ in eq.~(\ref{integral})) by one. 
This denotes the
efficient synchrotron cooling of the whole distribution.
We emphasize that the results
for  $\gamma \gtrsim \gcr$ should be taken cautiously into consideration, since  
the above solution is not valid for all $\gamma$ above $\gcr$; we remind that it 
was derived under the approximation $A\gamma^{\beta}>1$ or equivalently
\eqb
\gamma>\left(\frac{2^{\Gamma-1}(\Gamma+1)(\Gamma-1)\alpha_2}{Q_0\alpha_1\sigma_0 b^{\frac{\Gamma+1}{2}}}\right)^{2/\Gamma}.
\label{valid}
\eqe 
After having determined the form of the electron distribution we can then calculate the steady-state solutions
for $\gamma$-rays and soft photons. For $\epsilon\le \ecr$ the injected spectrum remains unaffected by automatic quenching, whereas
for $\epsilon>\ecr$, the solution $\nga(\epsilon)$ can be found by inserting first the second branch of $\nel$ into 
eq.~(\ref{soft}) and then use the derived expression for $\ns(x)$ into eq.~(\ref{gammaray}).
The results are summarized below
\eqb
\nga(\epsilon)&=&Q_0 \epsilon^{-\Gamma}, \ \emn \le \epsilon\le \ecr  \nonumber\\ 
\label{gamma-sol}
\phantom{}  \\  \nonumber
\nga(\epsilon)&=&\frac{Q_0 \epsilon^{-\Gamma}}{1+\frac{F}{\epsilon}\left(\left(\frac{2}{\epsilon}\right)
^{-\alpha}-\xmx^{-\alpha}\right)}, \  \ecr<\epsilon\le \emx
\eqe
where $\alpha=\Gamma/2+1$ and 
\eqb
F=\frac{8 b^{\frac{\Gamma+1}{2}} Q_0 
\sigma_0 \alpha_1 }{2^{\Gamma}(\Gamma-1)(\Gamma+2)\alpha_2}
\eqe
As the solution for $\nel$ is not valid for $\gamma \gtrsim \gcr$ (see eq.~(\ref{valid})),
the behaviour of $\nga$ close to the transition, i.e. for $\epsilon\gtrsim \ecr$, should also
be considered with caution. The validity range set by eq.~(\ref{valid}) can be translated into
terms of photon energies, i.e. $\epsilon>\epsilon_{\star}$, where
\eqb
\epsilon_{\star}= \left(\frac{2^{\frac{3\Gamma}{2}}(\Gamma-1)(\Gamma+2)\alpha_2}{4Q_0\sigma_0 b^{\frac{\Gamma+1}{2}}\alpha_1}\right)^{\frac{2}{\Gamma}}
\eqe
It can be easily verified that, for $\epsilon>\epsilon_{\star}$ the second term in the denominator of $\nga$ is larger than unity\footnote{The physical interpretation
of this condition is that the absorption term in the $\gamma$-ray photon equation is larger than the escape term. In this energy range, the photon distribution 
is determined mainly by the automatic quenching.}, which simplifies the functional dependence of $\nga$  on energy, i.e. 
$\nga\propto \epsilon^{-3\Gamma/2}$. Summarizing, we find that
the asymptotic photon index of a spontaneously quenched $\gamma$-ray spectrum is well defined and it is given by $3\Gamma/2$; 
this result is in complete agreement with the one derived numerically in PM11 -- see Fig.~9 therein. The spectral break in the case of 
automatic quenching depends, therefore, linearly on the photon index at injection, i.e. $\Delta \Gamma=\Gamma/2$.

\begin{figure}
\centering
 \includegraphics[width=8.5cm,height=7cm]{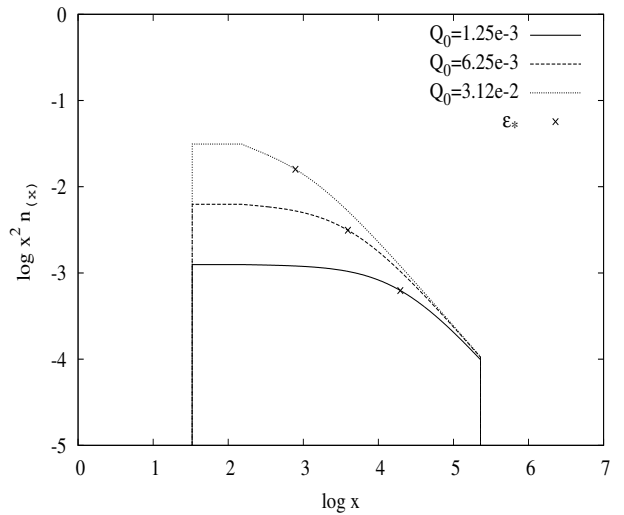}
\caption{Analytical solution of steady-state spontaneously quenched $\gamma$-ray spectra 
for different values of the injection
parameter $Q_0$ starting from $1.25\times10^{-3}$. Each value is increased by a factor of 5 over 
its previous one. $X$-symbols denote in each case the value of $\epsilon_{\star}$. 
Other parameters used are: $\Gamma=2$, $\emn=33$, $\emx=2.3\times10^5$, $B=40$ G and $R=3\times10^{16}$ cm.
The break energy is then $\ecr=160$.}
\label{fig1}
\end{figure}

\begin{figure}
\centering
 \includegraphics[width=8.5cm,height=7cm]{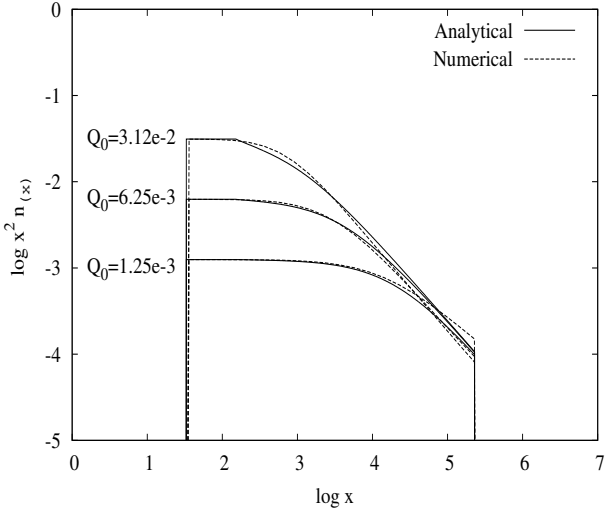}
\caption{Numerical (dashed lines) and analytical (solid lines) solutions for the  steady-state  $\gamma$-ray spectra.
The injection rate increases from bottom to top starting from $Q_0=1.25\times10^{-3}$. 
The parameters used are the same as in Fig.~\ref{fig1}. }
\label{fig2}
\end{figure}

Figure \ref{fig1} shows the $\gamma$-ray spectra given by eq.~(\ref{gamma-sol}) for $\Gamma=2$ along with
$\epsilon_{\star}$ ($X$-symbols) for different values of 
the injection parameter $Q_0$ that ensure the operation of spontaneous photon quenching. Our solution for $\epsilon>\ecr$
becomes progressively valid over a larger range of energies as $Q_0$ increases. Moreover, for the highest
value of $Q_0$, the photon index of the quenched part of the spectrum is $3$, i.e. it has obtained its asymptotic value 
defined by $3\Gamma/2$. 
The analytical solutions presented in Fig.~\ref{fig1} are compared to those
derived using the numerical code described in the following section and PM11. Solid and
dashed lines in Fig.~\ref{fig2} correspond to the analytical and numerical solutions respectively. 
The agreement between the two is better for larger values of the 
injection compactness, i.e. when the absorption term in the equation of $\gamma$-ray photons
becomes larger.

Before closing the present section and for reasons of completeness we make a short comment on our
choice of assuming continuous instead of catastrophic energy losses of electrons. We have also
derived the steady-state solution in the case of monoenergetic injection of 
$N$ $\gamma$-rays using catastrophic losses. However, the obtained results, when compared to the numerically derived ones,
were not reasonable, in the sense that no large spectral breaks were produced since $\gamma$-ray absorption was underestimated. 
 The main reason behind this disagreement with the numerically derived results is the approximation
(4). By assuming catastrophic losses  of the produced pairs we neglect
soft photons with $x<b \epsilon^2/4$, i.e. we artificially decrease the optical
depth for absorption of a $\gamma$-ray photon with certain energy $\epsilon$. Thus, although the assumption of catastrophic losses
proves to be suffiecient for the derivation of $\lcr$ (section 2.1) it proves to be too crude for more quantitative results.
\section{Numerical approach}
In this section we will present 
\begin{enumerate}[(i)]
 \item  the dependence of the critical injection 
compactness on various parameters, e.g. on the minimum and maximum energy of injected $\gamma$-rays,
as well as on on the $\gamma$-ray photon index, for a wide range of values.\\
\item the effects that the presence
of low-energy photons has on the automatic absorption of $\gamma$-rays.
\end{enumerate}
For a detailed study of the above 
a numerical treatment is required; as far as the first 
point is concerned, we have already shown that the analytical approach breaks down
e.g.  for sufficiently high values of the minimum
energy of injected $\gamma$-rays -- see section~2.1.

To numerically investigate the properties of quenching
one needs to solve again the system of eqs.~(\ref{4eqa})-(\ref{4eqb}), where the discretized photon number densities should
be replaced by their continuous functional form.
For completeness we have augmented it
to include more physical processes. 
As in the numerical code, there is no need to treat the time-evolution of
soft photons and $\gamma-$rays through separate
equations, the system can be written:

\eqb
{{\partial\nel(\gamma,t)}\over{\partial t}}+
{{\nel}\over{\teesc}}= \cal{Q}^{\rm e}_{\gamma\gamma}
+\cal{L}^{\rm e}_{\rm syn}
+\cal{L}^{\rm e}_{\rm ics}
\eqe
and
\eqb
{{\partial\nga(x,t)}\over{\partial t}}+
{{\nga}\over{\tgesc}}= \cal{L}_{\gamma\gamma}^\gamma
+\cal{Q}_{\rm syn}^\gamma
+\cal{Q}_{\rm ics}^\gamma
+\cal{L}_{\rm ssa}^\gamma
+\cal{Q}_{\rm inj}^\gamma,
\eqe
where $\nel$ and $\nga$ are the differential electron
and photon number densities, respectively, normalized as
in  section 2. Here we considered the following processes:
(i) Photon-photon
pair production, which acts as a source term for electrons
($\cal{Q}^{\rm e}_{\gamma\gamma}$) and a sink term for photons
($\cal{L}_{\gamma\gamma}^\gamma$); (ii) synchrotron radiation,
which acts as a loss term for electrons ($\cal{L}^{\rm e}_{\rm syn}$)
and a source term for photons ($\cal{Q}_{\rm syn}^\gamma$);
(iii) synchrotron self-absorption,
which acts as a loss term for photons (${\cal{L}}^{\gamma}_{\rm ssa}$)
and (iv) inverse Compton scattering, 
which acts as a loss term for electrons ($\cal{L}^{\rm e}_{\rm ics}$)
and a source term for photons ($\cal{Q}_{\rm ics}^\gamma$).
In addition to the above, we assume that $\gamma-$rays are
injected into the source through the term ($\cal{Q}_{\rm inj}^\gamma$).
The functional forms of the various rates have been presented
elsewhere -- \cite{mastkirk95,mastkirk97} and \cite{petromast09}. 
The photons are assumed to escape
the source in one crossing time, therefore $\tgesc=R/c$.
The electron physical escape timescale from the source
$\teesc$ is another free parameter which, however, is not
important in our case. Thus, we will fix it
at value $\teesc=\tgesc=R/c$. 
The final settings are the initial  conditions for the electron 
and photon number densities. Because we are investigating 
the spontaneous growth of pairs and their emitted synchrotron photons,
we assume that at $t=0$ there are no electrons in the source, so we set
$\nel(\gamma,0)=\eta \rightarrow 0$. Moreover, during the injection of photons 
in a certain $\gamma$-ray energy range it is important to keep 
the background photons used in the numerical code at a level 
as low as possible in order to avoid artificial growth of the instability.

\subsection{Critical $\gamma$-ray compactness for power-law injection}
The procedure we follow for the numerical determination of the critical
compactness is as follows: we start by injecting $\gamma$-rays at 
a rather low rate  in a specific energy range $(\emn,\emx)$, e.g.  $\linj\simeq 10^{-5}$, and then we
increase $\linj$ over its previous value by a factor of 0.2 in logarithm. 
 The increase of $\linj$ is directly related to the increase of the normalization $Q_0$ of the injection 
$\gamma$-ray spectrum; we remind that
$\linj = (Q_0/3)(\emn^{-\Gamma+2}-\emx^{\Gamma+2})/(\Gamma-2)$ for $\Gamma \ne 2$. 
For each value we allow the system
to reach a steady state and then we examine the shape of the $\gamma$-ray spectrum. 
We define as $\lcr$
that value of $\linj$ that causes the first deviation from the spectral shape at injection; for this reason,
we consider it as a rather strict limit. 

\begin{figure}
\begin{tabular}{l}
\includegraphics[width=9cm, height=7.5cm]{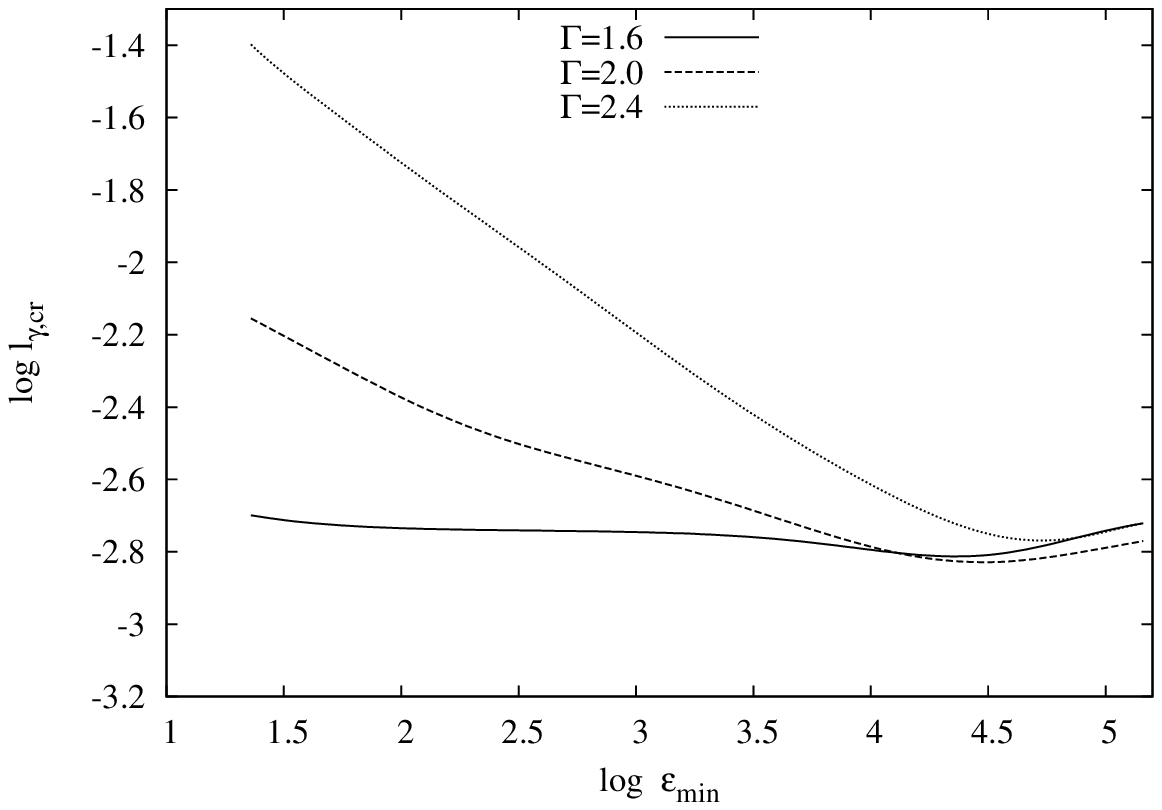} \\
\includegraphics[width=9cm, height=7.5cm]{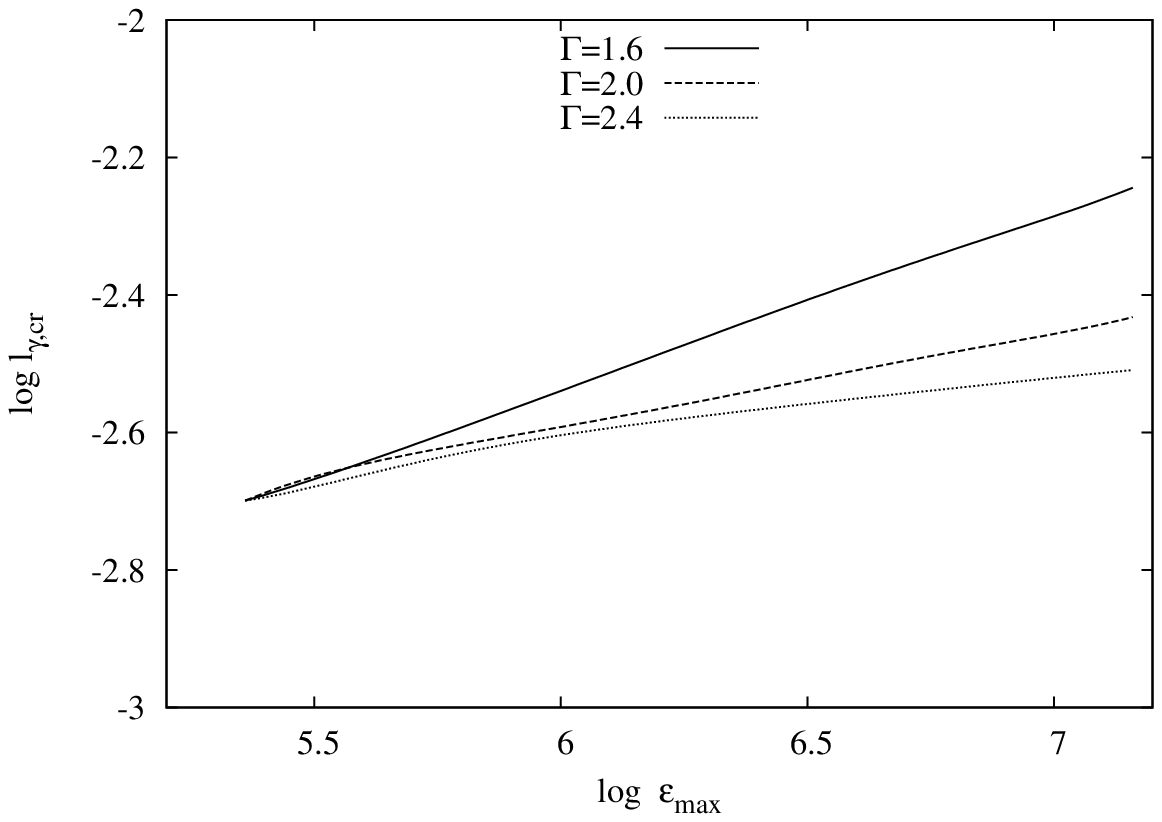}
\end{tabular}
\caption{Log-log diagram of the 
critical $\gamma$-ray compactness as a function of $\emn$ for constant $\emx=2.3\times10^5$  (top panel) 
and as 
a function of  $\emx$ for constant $\emn=1.4\times10^5$  (bottom panel)  for 
three photon indices marked on the plot. Other parameters used are: $B=40$ G
and $R=3 \times 10^{16}$ cm.}
\label{lcr-num}
\end{figure}

Figure \ref{lcr-num} shows $\lcr$ as a function of $\emn$ (top panel) or $\emx$ (bottom panel)
 for three different slopes of the injection
spectrum marked on the plot. Other parameters used are: $B=40$ G
and $R=3 \times 10^{16}$ cm. 
In the top panel, the maximum energy of the
power-law spectrum is fixed at $\emx=2.3\times10^5$, whereas
in the bottom panel the minimum energy is taken to be constant and equal
to $\emn=1.4\times10^5$.
For soft injection spectra, e.g. $\Gamma=2.4$, we find that $\lcr$ strongly depends on $\emn$.
The situation is exactly the opposite for hard $\gamma$-ray spectra -- see solid line in bottom panel of the same figure.
The critical compactness that we derive for $\emn \approx \emx \approx 2.5\times 10^5$ is $\lcr\approx 2\times10^{-3}$ -- see bottom panel of Fig.~\ref{lcr-num} --
and it corresponds to monoenergetic $\gamma$-ray
injection. Thus, it should be compared to 
$\lcr\approx 5\times10^{-4}$, which was derived analytically in PM11
for $\delta$-function injection at $\epsilon=2.5\times10^5$ -- see Fig.~2 therein.
The difference between the two results is not worrying, since the
analytical values in PM11 were about a factor of four lower than the accurate values that were derived numerically.

\subsection{Effects of a primary soft photon component}
The main difference between automatic $\gamma$-ray absorption
and the widely used photon-photon absorption on a preexisting photon field (`primary' photons), is that
in the first case no target field is initially present in the source. 
It is, as if the 
system finds its own equilibrium by self-producing the soft photons required for quenching
the extra $\gamma$-ray luminosity. 
In many physical scenarios, however, primary photons are present in the source, e.g. synchrotron radiation from
primary electrons, and therefore $\gamma$-rays are more likely to be absorbed on both primary and secondary
photon fields. Here arises the question whether or not the effects of spontaneous photon quenching
can be disentagled from those of linear absorption\footnote{Some discussion on the subject in the context
of differences in variability patterns can be found in PM12b.}.  

In the limiting cases where $\gamma$-rays are being absorbed on either primary or secondary
soft photons there is a straighforward relation between the photon index of the absorbed $\gamma$-ray spectrum
and that of photon target field:
\eqb
\label{index1}
\Gamma_{abs} & = & \Gamma+s-1, \ \textrm{`linear' quenching} \\ 
\Gamma_{abs} & = & \frac{3}{2}\Gamma,\ \textrm{`non-linear' quenching} 
\label{index2}
\eqe
where $s$ is the photon index of the primary photon distribution -- see Appendix for the derivation of
 $\Gamma_{\rm abs}$ in the first case.

\begin{figure}
\begin{tabular}{c}
 \includegraphics[width=8cm,height=7cm]{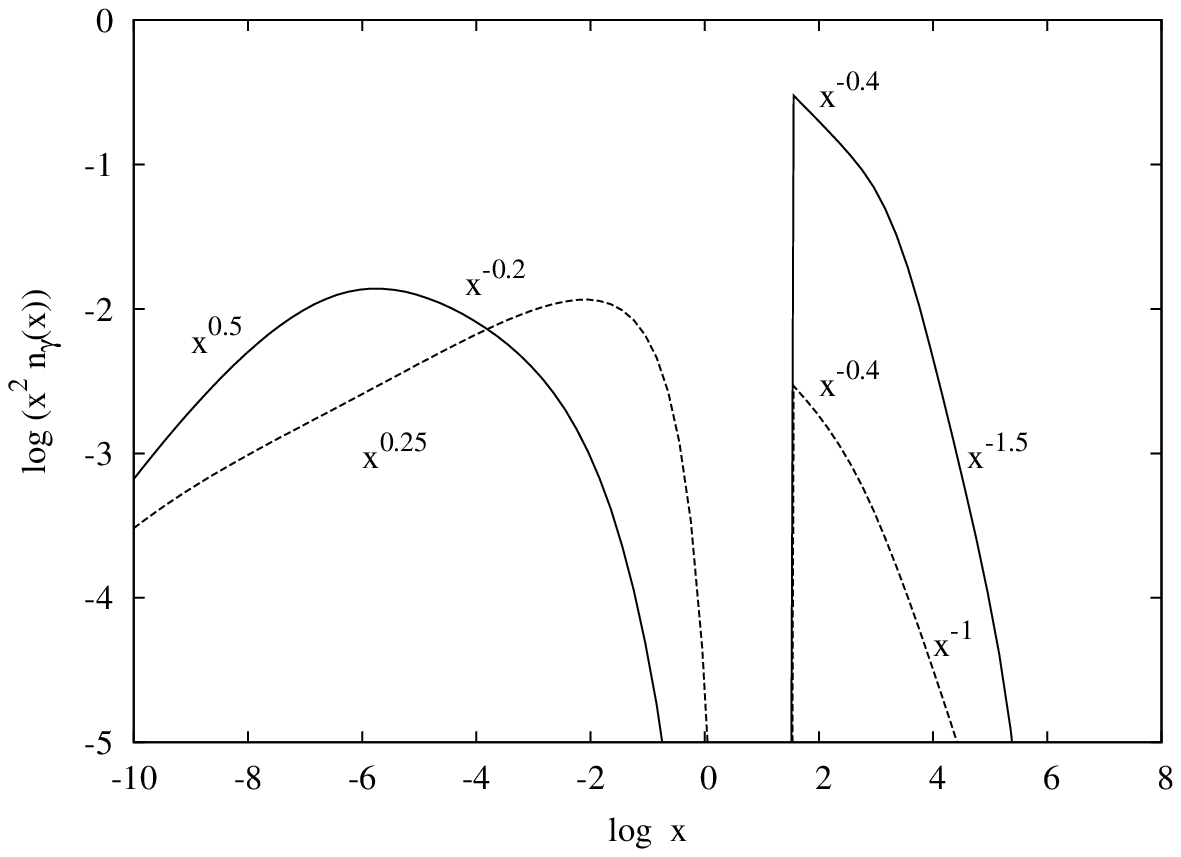} \\
 \includegraphics[width=8cm,height=7cm]{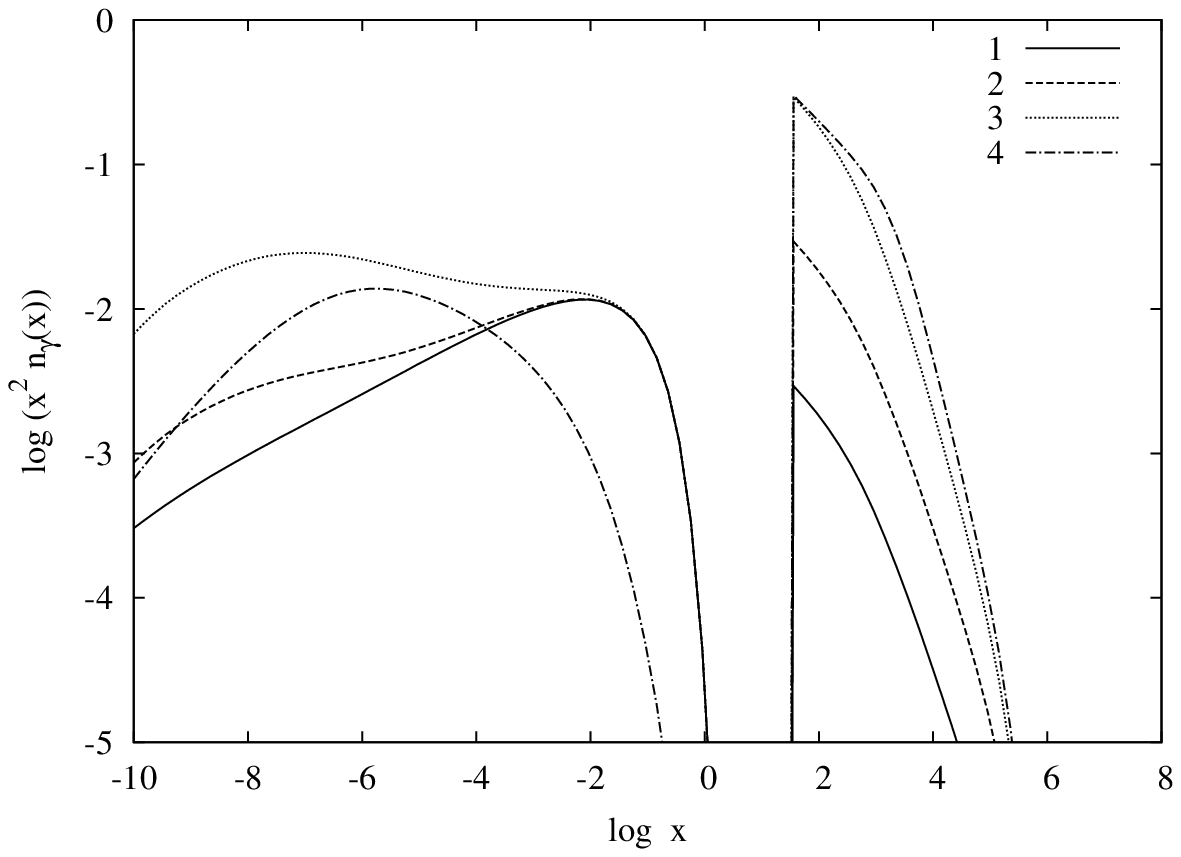}
\end{tabular}
\caption{Top panel: Comparison between two limiting cases where $\gamma$-ray photons
are absorbed only on synchrotron photons emitted by secondary pairs (solid line) or by primary electrons (dashed line). 
The injection compactnesses of $\gamma$-rays and primary electrons are $\linj=8\times10^{-1}, \leinj=10^{-7}$  (solid line) and 
$\linj=8\times10^{-3}, \leinj=3\times10^{-2}$ (dashed line) respectively.
Different parts of the spectra have different power-law dependencies that are marked on the plot.  Bottom panel:
Example of an indermediate case where $\gamma$-rays are being partially absorbed on synchrotron photons from secondaries.
The injection compactness of primary electrons is $\leinj=3\times10^{-2}$, while $\linj$ takes the following values:
$8\times10^{-3}$ (solid line), $8\times10^{-2}$ (dashed line) and $8\times10^{-1}$ (dotted line). The dash-dotted line
that is obtained for $\linj=8\times10^{-1}, \leinj=10^{-7}$ (same as solid line in top panel) is plotted for comparison reasons. 
Other parameters used for the plot are: $B=40$ G and $R=3\times10^{16}$ cm.} 
\label{abs2}
\end{figure}
Using the numerical code presented in section 3 we will first verify the above analytical relations and then
study intermediate cases.
For this, we assume a spherical region with size $R=3\times10^{16}$ cm containing a magnetic field 
$B=40$ G. VHE $\gamma$-rays with a photon index $\Gamma=2.4$ are injected into this
volume between energies $\emn=23$ and $\emx=2.3\times10^5$ with  compactness $\linj$ being
 a free parameter. Primary photons are produced via synchrotron radiation of a power-law electron distribution with
index $p=1.5$. 
The maximum Lorentz factor $\gmx$ and the electron injection compactness $\leinj$,  which is defined
as $\leinj = L_{\rm e}^{\rm inj} \sth / 4 \pi R \me c^3 $,  are also 
free parameters; here $L_{\rm e}^{\rm inj}$ is the total electron 
injection luminosity. The only processes we consider in the following examples are synchrotron emission, photon-photon
absorption and escape from the source. We note that the effect of inverse Compton scattering is negligible for
the magnetic field and electron energies assumed here.
The results regarding the two regimes are summarized in Fig.~\ref{abs2}. 
In the top panel we have used $\leinj=10^{-7}, \linj=8\times10^{-1}>>\lcr$  (solid line) and 
$\leinj=3 \times 10^{-2},\linj=8\times10^{-3}<<\lcr$ (dashed line), while we kept $\gmx=3.6\times10^5$ fixed.
For these parameter values the maximum energies of primary and secondary synchrotron photons are  
$10^{-1}$ and $1.3\times10^{-2}$ respectively. 
The slopes of different power-law segments in a $x^2\nga(x)$ plot are also shown.
The spectral break of the $\gamma$-ray spectrum differs
between the two regimes and the numerical results are in  agreement with those given by eqs.~(\ref{index1}) and (\ref{index2}).
In particular, we find that the absorbed $\gamma$-ray spectrum has a photon index $\Gamma_{\rm abs}=3.5$ which should be compared
to $3\Gamma/2 = 3.6$ for the `non-linear' quenching case.  In order to estimate the spectral change for the `linear' absorption case 
the photon index of the
soft photon distribution is required, which in this example is $s=(p/2)+1=1.75$. The expected value of $\Gamma_{\rm abs}$ is $3.15$ while
the derived one is $3$. We note also that the spectral shape
of the synchrotron component emitted by the produced pairs (solid line) is in agreement with that expected from
our solution for the electron distribution -- see eq.~(\ref{elec-sol}); emission from
lower energy electrons ($\nel \propto \gamma^{-2}$) results in  $\nga \propto x^{-3/2}$, while
from higher energy electrons ($\nel \propto \gamma^{-\Gamma-1}$) corresponds to  $\ns \propto x^{-\Gamma/2-1} = x^{-2.2}$.
In the bottom panel, along with the examples shown in the top panel (solid and dash-dotted lines), we plot
two intermediate cases
with a progressively higher $\gamma$-ray compactness, i.e. $\linj=8\times10^{-2}$ (dashed line) and $\linj=8\times10^{-1}$ (dotted line), while
we used the same $\leinj=3\times10^{-2}$. 
Already from $\linj=8\times10^{-2}$, which is still above the critical value,
 the contribution of photons produced via automatic $\gamma$-ray quenching is evident  
as a bump in the primary soft photon component. Notice also that the presence of primary soft photons enhances the automatic 
absorption of $\gamma$-rays for the same $\linj$ -- compare the dotted and dash-dotted lines.

\section{Implications on $\gamma$-ray emitting blazars}
\subsection{General remarks}
The mechanism of automatic photon quenching  sets an upper
limit to the intrinsic $\gamma$-ray luminosity of a compact source, and therefore,
can be applied on $\gamma$-ray emitting blazars for constraining 
physical quantities of their VHE emission region -- for more details
see PM11 and \cite{petromast12} -- hereafter PM12a. This can be relevant to 
recent observations that have revealed the
presence of very hard intrinsic
TeV $\gamma$-ray spectra of blazars, even when corrected with low EBL flux levels,
 e.g. 1ES 1101-232 \citep{aharonianetal06},
1ES 0229+200 \citep{aharonianetal07}. 
The SEDs of some hard $\gamma$-ray sources are very difficult
to be explained within one-zone emission models, and therefore, they
are often attributed to a second component, whose emission in longer wavelengths
is hidden by that of the first component (e.g.~\cite{costamante12}). 
Thus, the physical conditions of the component
 emitting in the TeV energy range can be chosen
quite arbitrarily only by demanding to fit the VHE part of the spectrum.
The presence of very hard TeV $\gamma$-ray spectra (typically $\Gamma_{\rm int} \simeq 1.5$)
implies that these sources cannot be spontaneously quenched, i.e. their intrinsic $\gamma$-ray 
luminosity should be less than the critical one. We note, however,
that there is an alternative scenario 
that employs the photon-photon absorption on internal soft photon fields
exactly for explaining the formation of very hard intrinsic $\gamma$-ray spectra \citep{aharonianetal08}.
\begin{table*}
 \caption{Photon indices of the GeV- and TeV-$\gamma$-ray spectrum
for the sources used in Fig.~\ref{index-obs}. Their redshift and the reference papers
are also listed.}
\label{table1}
\begin{threeparttable}[b]
\centering
\begin{tabular}{cccccc}
\hline 
\hline
\\
\phantom{} & Source & $\Gamma_{\rm TeV}$ &   $\Gamma_{\rm GeV}$\tnote{1} & $z$ & Reference \\
\hline\\
MAGIC  & 1ES 1215+303 & $2.96\pm0.14 $&  $2.0 \pm 0.2$&  0.130 & \cite{aleksicetal12} \\
\phantom{} & NGC 1275 & $4.1 \pm 0.7$ &  $2.17 \pm 0.05$ & 0.018 & \cite{aleksicetal12b}\\
\phantom{} & Mrk 421 & $2.48 \pm 0.03$ & $1.75\pm 0.03$ & 0.031 & \cite{abdoetal11} \\
\phantom{} & PKS 1222+21 & $3.75\pm0.27$ &$1.95\pm0.21$ &0.43& \cite{aleksicetal11} \\
\phantom{}& 3C 279 & $3.1\pm 1.1$ & $2.3\pm 0.1$ & 0.536& \cite{abdoetal09,aleksicetal11b}\\
\phantom{} & IC 310 & $2.00\pm0.14$ & $1.58\pm0.25$ & 0.019 & \cite{aleksicetal10}\\
\phantom{} & Mrk 501 & $2.51\pm0.05$ & $1.78\pm0.03$\tnote{2} &0.034 & \cite{mrk501_abdo11} \\
\hline 
\\
H.E.S.S. & 1ES 0414+009 & $3.45\pm0.25$ & $1.85\pm 0.18$ & 0.287 & \cite{hess12} \\
\phantom{} &    PKS 2005-489& $3.20\pm0.16$ & $1.79\pm 0.07$ &  0.071 & \cite{hess11} \\
\phantom{} & PKS 0447-439& $\mathbf{3.89\pm0.37}$ & $\mathbf{1.85\pm0.035}$ & 0.2\tnote{3} & \cite{abramowskietal13}\\
\hline 
\\
VERITAS & RGB J0710+591& $2.69\pm0.26$ & $1.46\pm0.17$ & 0.125& \cite{acciarietal10b}\\
\phantom{} & PKS 1424+240 & $3.8\pm0.5$ & $1.8\pm 0.07$ & unknown & \cite{acciarietal10a}\\
\phantom{}& RJX 0648.7+516 & $4.4\pm0.8$& $1.89\pm0.1$ & 0.179& \cite{aliuetal11}\\
\phantom{} & RBS 0413 & $3.18\pm 0.68$ & $1.57\pm0.12$ & 0.19 & \cite{aliuetal12}\\
\\
\hline
\end{tabular} 
\begin{tablenotes}
  \item[1] All observations in the GeV energy band 
are made with \textit{Fermi} satellite. For the
specific energy range where the photon index is calculated see corresponding reference.
\item[2] Photon index of the average GeV spectrum.
\item[3] Although the redshift of this source is still disputed, we use the value 0.2 that
is in agreement with most of the present estimates in literature. 
\end{tablenotes}
\end{threeparttable}
\end{table*}

By now there are several (quasi)simultaneous observations of blazars
in the GeV and TeV energy range, which clearly show that the $\gamma$-ray
spectrum cannot be fitted by a simple power-law over the whole GeV-TeV energy range, but
rather by a broken power-law. The change of the photon index
in many cases is large ($\Delta \Gamma \gtrsim 1$) and it 
cannot be explained using simple arguments, such as cooling breaks ($\Delta \Gamma=0.5 $).
In particular, for high-peaked blazars that are bright in the \textit{Fermi} energy band 
(e.g. Mrk421, PKS 2155-304, PKS 0447-439 etc.),
the peak of their SED seems to fall in the high-GeV energy part of the spectrum ($ \simeq 100$ GeV).
In these cases GeV and TeV emission correspond to parts of the spectrum below and above the high-energy hump respectively. Thus,
it is commonly considered that the VHE $\gamma$-ray spectra are intrinsically much softer (e.g. exponential cutoff effects) 
than the GeV--spectra \citep{costamante12}. 

Here, however, we investigate another explanation of $\gamma$-ray spectral breaks.
In our framework, the \textit{injection} $\gamma$-ray spectrum
is described by a single power-law from GeV up to TeV energies, and the spectrum of the \textit{escaping}
$\gamma$-ray radiation is modified due to internal spontaneous photon absorption. 
We remind that the instability of automatic photon quenching offers
an alternative mechanism for producing intrinsic
broken power-law spectra (see  section 2).

\begin{figure}
\centering
 \includegraphics[width=9cm, height=8cm]{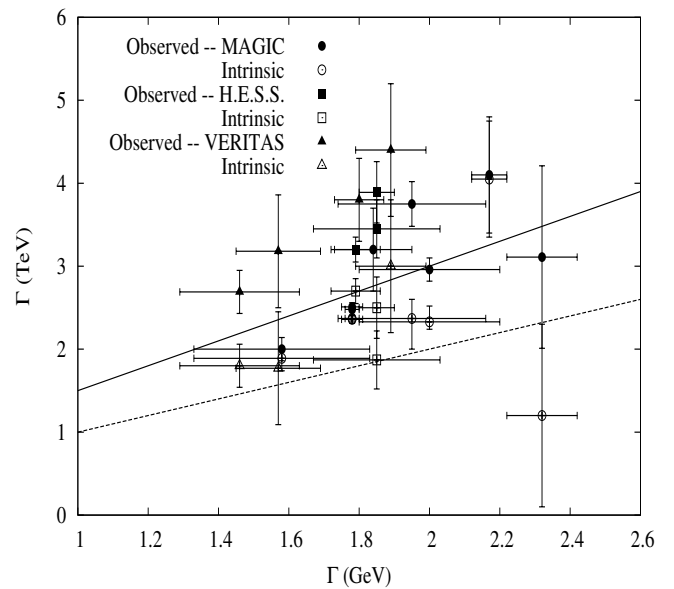}
\caption{Photon index of the TeV spectrum versus the 
one of the GeV spectrum for several $\gamma$-ray emitting blazars.}
\label{index-obs}
\end{figure}

A plot of the photon index in the GeV energy band (as measured by the \textit{Fermi} satellite)
versus the one in the TeV energy band (as measured by MAGIC and H.E.S.S. telescopes)
is shown in Fig.~\ref{index-obs}. 
The sources used for this plot are listed in Table~\ref{table1} along with the observed values of
the photon indices and the reference paper.
Filled and open symbols show the photon indices of the observed and of the corrected for EBL absorption
VHE spectra. We note that in all cases we used the model C by \cite{finkeetal10} for the EBL correction; 
in this model, the EBL flux from UV to the near -- IR is also similar to that of \cite{dominguezetal11}.
Different symbols, in particular circles and squares, denote TeV observations made by MAGIC and H.E.S.S. respectively.
The solid line represents the relation between the photon indices that is
expected by spontaneous photon quenching, i.e. $\Gamma_{\rm TeV}=3\Gamma_{\rm GeV}/2$,
whereas the dashed line ($\Gamma_{\rm TeV}=\Gamma_{\rm GeV}$) is plotted only for guiding the eye.
Some of the data points within their error bars are compatible with the theoretical predictions. For the purposes of
the present work, the choice of a particular EBL model does not affect significantly our results. For example, the EBL model used here predicts
slightly higher optical depth than the one of \cite{franceschinietal08} for $E_{\gamma} \lesssim$ 5 TeV (10 TeV) for $z=0.6$ (0.1) respectively. Thus,
correction of the VHE spectra with the EBL model of \cite{franceschinietal08}  would result
 in slightly softer intrinsic VHE spectra, i.e. some of the open
symbols in Fig.~\ref{index-obs} would move upwards in the vertical direction.
 Moreover, the fact that the analysis of the {\sl Fermi} and MAGIC/H.E.S.S./VERITAS data has not been made over the
same energy intervals for all sources listed in Table~\ref{table1} makes difficult the derivation of any conclusions regarding
the statistical properties of this sample. Non-simultaneity of GeV and TeV observations in some cases, e.g. NGC 1275, makes also
any coherent comparison difficult. 
For these reasons, this type of plot could be used only as a \textit{first indicator} for searching among sources that could be explained
by the mechanism of automatic quenching.

In these cases, there are five additional features that can be tested observationally:
\begin{enumerate}[(i)]
 \item Automatic photon quenching is a radiative instability that redistributes the energy
within a photon population. The absorbed $\gamma$-ray luminosity appears, therefore,
in the lower part of the multiwavelength (MW) spectrum--usually in the X-ray regime. 
\item The spectral break in the $\gamma$-ray spectrum of blazars that 
have low X-ray emission with respect to that of VHE $\gamma$-rays can, in general, not be
attributed to automatic photon absorption. 
\item If the VHE $\gamma$-ray spectrum is spontaneously absorbed, there is
a straighforward relation 
between the photon indices of the absorbed part of the $\gamma$-ray
spectrum and that of the soft photon component.
 In Section 2.3 we have shown that $\Gamma_{\rm abs}=3\Gamma/2$, where $\Gamma$ 
is the photon index of the $\gamma$-ray spectrum at injection. The steady state
electron distribution due to pair production is $\nel(\gamma)\propto \gamma^{-\Gamma-1}$ -- see eq.~(\ref{elec-sol}) --
and the corresponding photon index of the synchrotron spectrum is given by $\Gamma_{\rm soft}=\Gamma/2+1$. 
Thus, the relation between the two photon indices is  $\Gamma_{\rm abs} = 3(\Gamma_{\rm soft}-1)$ -- see also the numerical
example in the top panel of Fig.~\ref{abs2}.

\item Strong correlation between the soft component of the MW spectrum
and the unabsorbed part of the $\gamma$-ray spectrum
is to be expected in case where the intrinsic $\gamma$-ray
luminosity varies.
\item An increase of the intrinsic $\gamma$-ray compactness is accompanied
by a shift of the break energy towards lower energies.
\end{enumerate}
As far as the first observational
prediction is concerned, one can show that the maximum energy of
the soft component produced by automatic photon quenching, falls,
for reasonable parameter values,  in the X-ray regime.
First, the `feedback' criterion for automatic quenching must
be satisfied, at least, by $\gamma$-ray photons having the 
maximum energy (see also section 2).  This is written as
\eqb
B > B_{\rm q} \equiv (4\times10^{-5})\ \delta^{3} (1+z)^{-3} \left(E_{\rm max,12}^{\rm obs}\right)^{-3} \ \textrm{in G},
\label{bq}
\eqe
where we have used the 
observed quantities instead of those measured in the comoving frame of the blob that has a Doppler factor $\delta$.
From this point on and in what follows the 
convention $E_{X}\equiv E/10^{X}$ in eV will be adopted for photon energies, unless stated
otherwise\footnote{Capital and small initial letters are used  for differentiating
 between photon energies with and without dimensions respectively.}.
In section 2 we have shown that a spontaneously quenched $\gamma$-ray spectrum
shows a break at the energy $\ecr=8\Bcr/(B\emx^2)$. 
Combining this expression with the fact
that the observed break energy $E_{\rm br}^{\rm obs}$
usually lies in the GeV energy band, we derive a second relation between the magnetic
field and the Doppler factor of the blob 
\eqb
B_{\rm br}=(4\times10^{-2})  \delta^3 (1+z)^{-3} \left(E_{\rm max,12}^{\rm obs}\right)^{-2} 
 \left(E_{\rm br,9}^{\rm obs}\right)^{-1} \textrm{in G},
\label{Bbr}
\eqe
Since the break energy is by definition smaller 
than the maximum one, the magnetic field $B_{\rm br}$
always satisfies inequality (\ref{bq}). 
We note that the magnetic field required is, generally, strong.
Even for a small value of the Doppler factor, e.g. $\delta=10$, 
one needs $B\simeq 40$ G -- see also PM12a for a related discussion.
Thus, spontaneously quenched $\gamma$-ray spectra cannot
operate in the context of one-zone leptonic models, such as synchrotron-self Compton (SSC) that usually requires weak magnetic fields (e.g. \cite{boettcher09}).
Finally, the observed maximum energy of the produced soft photons  
is given by
\eqb
E_{\rm s,max}^{\rm obs}=0.5 \ \delta^2 (1+z)^{-2} \left(E_{\rm br,9}^{\rm obs}\right)^{-1} \ \textrm{in keV}
\eqe
where we have used the magnetic field strength given by eq.~(\ref{Bbr}). 

\begin{figure}
\centering
 \includegraphics[width=9cm, height=7cm]{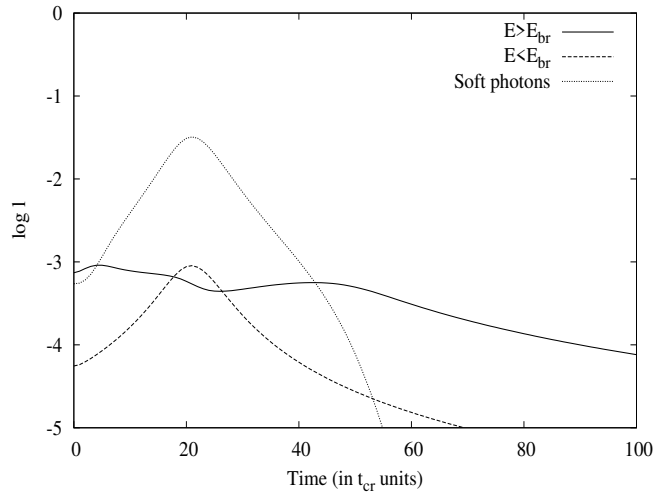}
\caption{Plot of photon compactness as a function of time for various components: absorbed and unabsorbed $\gamma$-rays are plotted with
solid and dashed lines respectively, while soft photons are shown with a dotted line.
Parameters used for this plot: $\emx=2.3\times10^5$, $\emn=23$, $\Gamma=1.5$, $B=40$ G, $R=3\times10^{16}$ cm,
$\linj=1.7\times10^{-3}\gtrsim \lcr$, $G=10$ and $\tau_{\rm c}=20$.}
\label{quench-var}
\end{figure}

\begin{figure}
\centering
 \includegraphics[width=9cm, height=7cm]{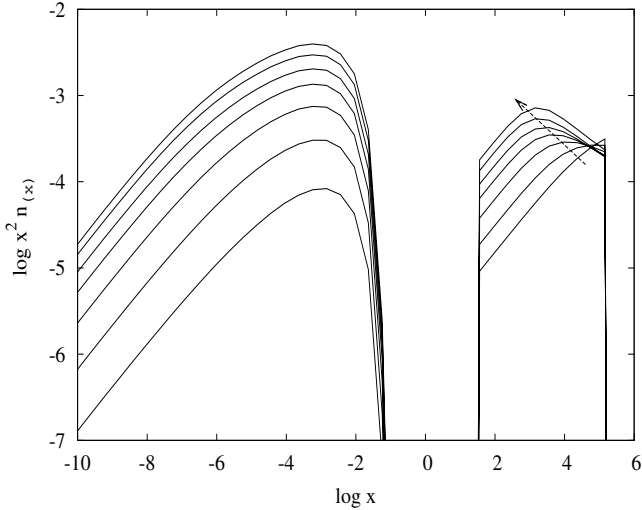}
\caption{Snapshots of MW spectra obtained for a variable $\gamma$-ray injection
compactness. The type of variation and the parameter values are the same as in Fig.~\ref{quench-var}. The arrow
shows the `hard-to-soft' evolution of the $\gamma$-ray spectrum break energy.}
\label{quench-var-spec}
\end{figure}

If a $\gamma$-ray emitting blazar happens to be a spontaneously quenched source,
then one can make a strong prediction about 
the flux correlation between 
soft (usually X-ray) photons, 
the unabsorbed part of the $\gamma$-ray
 spectrum (GeV energy band) and the absorbed one (typically TeV energy band).
Increase of the intrinsic $\gamma$-ray compactness amplifies, in general, 
the absorption of VHE $\gamma$-rays, which leads to an increase
of the soft photon component. The number of photon targets  for the $\gamma$-rays 
is then increased, which further sustains the non-linear
loop of photon-photon absorption. The part of the $\gamma$-ray spectrum 
that is not affected by automatic quenching
follows exactly the variations of the injection compactness, whereas the spontaneously quenched
varies in the inverse way.
The above are exemplified in Fig.~\ref{quench-var}. 
First, we allowed the system to reach a steady state -- for the parameters used see caption of Fig.~\ref{quench-var}. 
Then, we imposed  on the injected $\gamma$-ray compactness a Lorentzian variation
\eqb
\linj(\tau) = \left(\linj\right)_{0}\frac{F_{\rm L}(\tau)}{F_{\rm L}(0)}
\eqe
where $\tau$ is the comoving time in units of $\tcr$ and 
\eqb
F_{\rm L}(\tau)=\frac{\frac{G}{2\pi}}{(\tau-\tau_{\rm c})^2+\left(\frac{G}{2\pi}\right)^2}
\eqe
where $\tau_{\rm c}, G$ are free parameters that control the  position of the maximum
and the width at half maximum respectively.
Then, we calculated the photon compactness of the soft component ($x\le b\emx^2/4$),
of the unabsorbed ($\emn\le\epsilon\le 10\emn \approx \ecr$) and spontaneously quenched ($0.1\emx\le\epsilon\le \emx$) $\gamma$-rays
and plotted them as a function of time in Fig.~\ref{quench-var}.
The evolution of the break energy of the spectrum from high to lower energies can be seen in Fig.~\ref{quench-var-spec}, 
where snapshots of MW spectra
are plotted.

\subsection{Application to blazar PKS 0447-439}

PKS 0447-439 is a bright blazar that has been recently detected at high energy 
\citep{abdoetal09b} and VHE \citep{abramowskietal13} $\gamma$-rays. 
Although the redshift of the source is still disputable (see e.g. \cite{prandinietal12, pitaetal12} 
for different estimations and discussion),
in the following we will adopt the value $z=0.2$. Furthermore, a higher redshift, 
that is equivalent to a larger optical depth for $\gamma \gamma$ absorption, would imply a 
harder intrinsic VHE spectrum and therefore a smaller value of 
$\Delta \Gamma \equiv \Gamma_{\rm TeV}-\Gamma_{\rm GeV}$.
 However, taking into account the errorbars of the photon indices, a fit can still be achieved for
a range of redshift values ($0.1<z<0.24$).
We have focused on PKS 0447-439 since it satisfies most of the conditions 
that were presented in the previous section. In particular:
\begin{itemize}
 \item The photon indices in the GeV- and TeV- energy range are $\Gamma_{\rm GeV}= 1.85 \pm 0.05$
and $\Gamma_{\rm TeV} \approx 2.5 \pm 0.37$ respectively\footnote{The VHE photon index is calculated after correcting
for EBL absortpion for $z=0.2$ and using the model by \cite{finkeetal10}.}. This is in agreement with what we have shown, i.e.
 a $\gamma$-ray spectrum with $\Gamma=1.8$
can steepen up to $\Gamma=2.7$ due to spontaneous quenching. 
\item The X-ray luminosity is less than 
the $\gamma$-ray one but of the same order of magnitude. 
\item An anticorrelated variability between VHE $\gamma$-rays and X-rays  may be suggested, 
although the number of data points is small and therefore still inconclusive --  see Fig.~5 in \cite{abramowskietal13}. 
\end{itemize}

\begin{figure}
\centering
 \includegraphics[width=9cm, height=7cm]{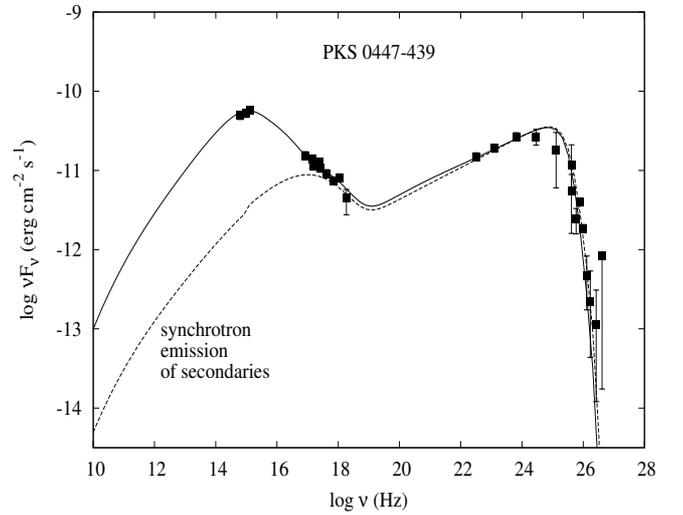}
\caption{Multiwavelength spectrum of PKS 0447-439 during the period  November 2009-January 2010. 
Filled squares represent the \textit{Swift}/UVOT,  \textit{Swift}/XRT,
\textit{Fermi} and H.E.S.S. data from low to high energies respectively. Solid and dashed lines
show the SEDs with and without the injection of primary electrons.
 Our model SEDs are corrected for EBL absorption asumming $z=0.2$ and using model C of \cite{finkeetal10}.}
\label{fit1}
\end{figure}

As a first attempt, we did not specify the production mechanism of $\gamma$-rays.
Instead, we have assumed that they are being injected into the emission region with a rate given by
\eqb
Q_\gamma=Q_{0} \epsilon^{-\Gamma}H(\epsilon-\emn)H(\emx-\epsilon),
\eqe
where $Q_0$ is a normalization constant that is related to the injection
$\gamma$-ray compactness as 
\eqb
Q_0 = \left\{ 
\begin{array}{l}
3\linj \ln\left(\frac{\emx}{\emn}\right)^{-1}, \ \textrm{if} \ \Gamma=2 \\ 
\phantom{} \\
3\linj (\Gamma-2)\left(\emn^{-\Gamma+2}-\emx^{-\Gamma+2}\right)^{-1}, \ \textrm{if} \ \Gamma \ne 2.
\end{array}
\right.
\label{qo-l}
\eqe
We have also included the injection of  primary electrons at a rate 
\eqb
Q_{\rm inj}^{\rm e} = Q_{0,e} \gamma^{-p} H(\gamma-\gmn)H(\gmx-\gamma),
\eqe
 where $Q_{0,e}$ is related
to the electron injection compactness in the same way as in eq.~(\ref{qo-l}). 
\begin{table}
\centering
\caption{Parameters for multiwavelength fit to the observations of PKS 0447-439 
during the period November 2009-January 2010 (see Fig.~\ref{fit1}).}
\label{table2}
\begin{threeparttable}[b]
\begin{tabular}{c  c}
 \hline 
Parameter symbol  & \\
\hline 
\hline
 $R$ (cm) & $8 \times 10^{15}$  \\
 B (G) &  20 \\
$u_{\rm B}$  (erg cm$^{-3}$) & 16 \\
 $\delta$ & 11  \\
\hline
  $\emx \ (\me c^2)$ & $10^5$ \\
$\emn \ (\me c^2)$ & $10^{-6}$\\ 
  $\Gamma$ &  1.8 \\
  $\linj$ & $\mathbf 7.2\times10^{-3} $\\
 $\lcr$ & $2\times10^{-3}$ \\
\hline
 $\gmx$  & $ \mathbf 8\times10^3$\\
$\gmn$ & $8\times10^2$ \\
 $p$ &   2.5 \\
$\leinj$ &  $2\times10^{-3}$ \\
\hline
\end{tabular}
 \end{threeparttable}
\end{table}
The parameters of the fit shown in Fig.~\ref{fit1} can be found in Table~\ref{table2}.

Figure \ref{fit1} shows that the synchrotron emission of secondary electron/positron pairs 
explains the X-ray emission, while the synchrotron emission from primary leptons
is required for fitting the optical data. In this example, 
primary soft photons, i.e. photons that are not produced by automatic quenching of $\gamma$-rays,
are not targets for photon-photon absorption  due to their low energies, and therefore, `linear' absorption does not
interfere with the non-linear one. We note that  inverse Compton scattering was taken also into account; both primary and secondary soft photon
fields were used in the scatterings.
As a second step, one could attribute $\gamma$-ray emission to 
a particular production mechanism, e.g. to 
synchrotron radiation from
relativistic protons, since the choice of a relatively large value of the magnetic field
makes the application of automatic photon quenching more relevant to 
hadronic emission models; a detailed hadronic modelling of the source lies, however,
outside from the scope of the present paper.

\section{Discussion}


A very interesting, yet largely unnoticed, 
property of $\gamma$-ray radiation transfer is the 
presence of an upper limit at 
the production rate per volume of $\gamma$-ray photons.
If the $\gamma$-ray compactness at injection exceeds this critical value,
then soft photons are produced spontaneously in the source, serve as targets for high-energy photons
and absorb the `excessive' $\gamma$-ray luminosity. 
Thus, soft photons act as a
thermostat and appear 
irrespective of the $\gamma$-ray production mechanism.
These ideas were put forward in SK07, who coined the term
`automatic photon quenching' to describe this non-linear
mechanism, and were expanded by PM11 and PM12b.  
The present paper continues the exploration of 
automatic quenching in the case where 
$\gamma$-rays are injected with a power-law distribution. 
Therefore, this can 
be considered as a continuation of an earlier work (PM11),
in which the quenching in the case of monoenergetic 
$\gamma$-ray injection was studied.

In  section 2 we have derived an analytical expression
of the critical compactness that is required 
for an injected power-law $\gamma$-ray spectrum to be quenched.
A series of approximations/assumptions, which were presented
in detail in the same section, were necessary for the above derivation.
 In particular, the `catastrophic losses' approximation for electrons proved to be crude enough and made our analytical
results valid in a particular parameter range; we have commented
on that through an indicative example, where the analytically and 
numerically derived values of the critical compactness were compared.
In cases where automatic photon quenching applies,
we have calculated the steady-state $\gamma$-ray spectra
and shown that spontaneous photon absorption produces
 a break of $\Delta \Gamma = \Gamma/2$ between the unabsorbed and the absorbed parts
 of the injected power-law; here $\Gamma$ is the slope of the injected $\gamma-$rays.

In section 3 we have implemented a numerical code that 
solves the full radiative transfer problem 
inside a spherical volume, in order to 
derive the critical $\gamma$-ray compactness
for a wide range of parameter values, e.g. for various
photon indices, as well as for different 
minimum and maximum energies of the $\gamma$-ray spectrum. 
 We have also examined the effects 
that a primary soft photon component would have on the absorption
of $\gamma$-rays. We found that these depend on both the compactness
 and the spectral shape of the external component.

In \S4 we have examined the implications of quenching on $\gamma$-ray
emitting blazars. We have also given a set of
crireria that can be used in order to deduce, 
observationally, whether a $\gamma$-ray 
blazar is spontaneously quenched or not. 
We note that the relevant information
is imprinted both on the SED and on the variability patterns
of the source. We also point out that blazar PKS 0477-439 meets
several of the criteria and there is a distinct possibility
that its high-energy spectrum is quenched, while the X-rays
are produced from the reprocession of $\gamma$-rays. 

In the present paper we have intentionally avoided pinpointing 
a specific mechanism 
for $\gamma$-ray production. However, the range of parameters used
and especially the choice of a rather large magnetic field value
(of the order of a few Gauss), imply that these ideas
are more effectively applied to hadronic models. In this case 
$\gamma$-rays could be produced either by proton-synchrotron 
radiation or by pion production \citep{MPD13}.

Automatic quenching might have some far-reaching implications for
$\gamma$-ray blazars. If there is evidence that it operates 
at some level, then part (or all) of the UV/X-ray component 
should be reprocessed $\gamma$-ray emission, i.e. there is no need for 
a primary component to produce all of the observed soft radiation. 
It also predicts that, for AGN related parameters, breaks in the $\gamma$-ray
spectra should appear in the high GeV --  low TeV regime. Moreover, a `hard-to-soft'
evolution of the spectral break is expected whenever the injected  $\gamma$-ray flux increases.
Therefore, future observations, especially with CTA \citep{Soletal13},
could prove decisive in detecting the presence of such spectral breaks. 
If, on the other hand, the sources do not show signs of spontaneous quenching, then some
interesting constraints apply to the source parameters, as this relates the 
$\gamma$-ray luminosity to the size of the source, the magnetic field strength
and the Doppler factor -- see PM12a for such an application on 
 quasar 3C 279. These constraints can be quite severe, especially
if the source undergoes strong flaring episodes and the absence of quenching
could only mean either a very large value of the Doppler factor or 
a low magnetic field in the production region. 
Both aspects have strong 
implications for the physical conditions prevailing in the emitting regions
of $\gamma$-ray blazars.

\begin{acknowledgements}
 We would like to thank the anonymous referee for useful comments/suggestions
on the manuscript and Dimitri Liggri for helping us compile
Table~\ref{table1}.
This research has been co-financed by the European Union 
(European Social Fund – ESF) and Greek national funds through the 
Operational Program "Education and Lifelong 
Learning" of the National Strategic Reference Framework (NSRF) - Research Funding Program: Heracleitus II. 
Investing in knowledge society through the European Social Fund. 
\end{acknowledgements}
\appendix
\section{$\gamma$-ray spectral break due to absorption on primary soft photons}
We derive the spectral break of a $\gamma$-ray spectrum due to the absorption
of a primary soft photon component ($\next$) that is present in the emission region. In this
case, the absorption of $\gamma$-rays is  a `linear' process in contrast to automatic quenching.
We treat the target photon population as a photon `tank', in the sense that $\next$ does not evolve with time.
Thus, the steady-state $\gamma$-ray photon spectrum is derived by solving
the same set of equations as those presented in section 2.3, with the operator 
$\cal{L}_{\gamma \gamma}^{\gamma}$ (and $Q_{\gamma \gamma}^{\rm e}$) being sligthly different:
\eqb
\cal{L}_{\gamma \gamma}^{\gamma} & = & -\frac{\nga(\epsilon)\sigma_0}{\epsilon}\int_{2/\epsilon}^{X_{\rm M}}  \textrm{d}x \ x^{-1}
 \left(\ns(x)+\next(x) \right)
\eqe
where $X_{\rm M}=\max[\xmx,x_0]$ and $\xmx, x_0$ are the maximum energies of the secondary and primary soft photons respectively. The above expression represents
the most general physical case, where $\gamma$-rays are being absorbed by both primary and secondary soft photons. 
One can distinguish between two regimes:
\begin{enumerate}
 \item spontaneously or `non-linear' $\gamma$-ray absorption, if  $\next(x)<<\ns(x)$ \\
\item `linear' $\gamma$-ray absorption, if  $\next(x)>>\ns(x)$.
\end{enumerate}
Here we focus on the second regime, where after following the same steps 
as those described in section 2.3, we find 
the steady-state 
$\gamma$-ray distribution
\eqb
\nga(\epsilon) \approx \frac{Q_{\gamma}(\epsilon)}{1+\frac{\sigma_0 n_0}{\epsilon}\int_{2/\epsilon}^{x_0}\textrm{d}x \ x^{-s-1}},
\eqe
where $n_0$ is the normalization of the primary soft photon distribution. Since we
are interested in calculating the photon index of the absorbed part of the spectrum it is sufficient
to look at the asymptotic expression of $\nga(\epsilon)$, 
\eqb
\nga(\epsilon) \rightarrow \epsilon^{-\Gamma-s+1}
\eqe
which 
is obtained 
for $\epsilon^{s-1}>> 2^{s} s/\sigma_0 n_0$.
Thus, in this regime the spectral break is given by
\eqb
\Delta \Gamma = s-1.
\eqe
\bibliographystyle{aa}
\bibliography{mardaf13}
\end{document}